\documentclass[twoside,12pt]{article}
\usepackage{epsfig}
\usepackage{graphics}

\def\Journal#1#2#3#4{{#1} {#2} (#4) #3 }

\def\PRC{{\em Phys. Rev.} C}

\newcommand{\be}{\begin{equation}}
\newcommand{\ee}{\end{equation}}
\newcommand{\bea}{\begin{eqnarray}}
\newcommand{\eea}{\end{eqnarray}}

\begin{document}

\title{  Electron Capture in $^{163}Ho$, Overlap plus \\ 
Exchange Corrections and the Neutrino Mass. \\} 
\author{Amand Faessler$^{1}$, Loredana Gastaldo$^{2}$,
F. \v{S}imkovic$^3$, \\
\\
$^1$Institute of Theoretical Physics, University of Tuebingen, Germany\\
$^2$Kirchhoff Institut f\"ur
 Physik, University of Heidelberg, Germany\\
$^3$JINR, 141980 Dubna, Moscow Region, Russia and \\
Comenius University, Physics Dept., \ $SK-842 15$ \  Bratislava, 
Slovakia.}
\maketitle
\begin{abstract} Holmium 163 offers perhaps the best chance 
to determine the neutrino mass by electron capture. 
This contribution treats the electron capture in $^{163}Holmium $ 
completely relativistic for the overlap and exchange corrections 
and the description of the bolometer Spectrum.  
The theoretical expressions are derived 
consistently in second quantization with the help of Wick's 
theorem assuming single Slater determinants for the 
initial Ho and the final Dy atoms with holes 
in the final $ns_{1/2}$ and $ np_{1/2}$ states. 
One needs no hand waving arguments 
to derive the exchange terms.   It seems, that for the first time the multiplicity 
of electrons in the orbital overlaps are included in the numerical treatment. 
Electron capture $ e^- + p \rightarrow n + \nu_e $ is 
proportional to the probability to find the captured electron in the parent atom 
at the nucleus. Non-relativistically this is only possible for $ns_{1/2}$ 
electron states. Relativistically 
also $p_{1/2}$ electrons have a probability due to the 
lower part of the relativistic electron spinor, 
which does not disappear at the origin. Moreover relativistic effects 
increase by contraction the electron probability 
at the nucleus. Capture from other states are suppressed. However they can be 
allowed with smaller intensity due to finite nuclear size. 
These probabilities are at least three 
orders smaller than the electron capture from $3s_{1/2}$ 
and $3p_{1/2}$ states. 
The purpose of this work is to give a consistent
 relativistic formulation and treatment of the overlap 
and exchange corrections for electron capture in  $^{163}_{67}Ho$ to excited atomic states 
in $^{163}_{66}Dy $ and  to show the influence 
of the different configurations in the final Dy states. The overlap and exchange corrections 
are essential for the calorimetric spectrum of the 
deexcitation of the hole states in Dysprosium. The slope 
of the upper end of the spectrum,   
which contains the information on the electron neutrino mass, is different. 
In addition the effect of 
the finite energy resolution on the spectrum and on the 
determination of the neutrino mass is studied. 
The neutrino mass must finally be determined by maximum likelihood 
methods to fit the  theoretical 
spectra at the upper end near the Q value varying the neutrino mass, 
the Q-value and probably also the energy resolution, because at 
the moment Q and the energy resolution are not known to the accuracy needed. 

\end{abstract}

\section{Introduction}

Neutrino oscillations give the differences of the mass eigenvalue $m_i \ (i \ = \ 1,\ 2,3)$ 
squared, 
but they cant determine the absolute value of the neutrino mass eigenstates. The Tritium beta decay gives 
for direct determinations presently the best upper limit of about 2.2 [eV] 
for the anti-electron neutrino mass \cite{Mainz, Lobashev, KATRIN}. In the future one expects, 
that this value will be improved by the KATRIN 
experiment \cite{ KATRIN}. The electron Majorana neutrino mass 
can be determined by the neutrinoless double beta 
decay \cite{GERDA, Fang2, Simko}. The electron 
neutrino mass can  also be obtained by electron capture on a proton in nuclei
 \cite{Gastaldo1, Gastaldo2, Blaum, Fang, Krivo, Gastaldo5}.     

\be
 p + e^- \rightarrow n + \nu_e  \label{reaction}
\ee

For electron capture  the decay  $^{163}_{67}Ho_{96} \rightarrow ^{163}_{66}Dy_{97} $ 
seems to be the most promising case 
due to the small Q-value between 2.3 [keV] \cite{Anderson} 
and 2.8 [keV] \cite{Blaum, Gatti, Ra} with a recommended value of 
$Q \  = \ (2.55 \ \pm \ 0.016) \ keV$  \cite{Wang}. 

\be
 Q = M(^{163}Ho) -M(^{163}Dy) \approx 2.5 keV \label{Q}
\ee

In the electron capture process one can distinguish two steps \cite{de}. 
First the electron is captured in the state $|k>$ in $^{163}Ho$, an electron neutrino 
is emitted with energy $E_\nu$  and an
excited state is formed in $ ^{163}Dy$  with a hole in the electron orbital $ |f'>$  
with the same quantum numbers as $|f>$ in $^{163}Ho$. 
A calorimetric measurement "detects" the total energy of the second 
step plus the nuclear recoil occurring in the first stage, which is of the order of meV 
and will be neglected in the following discussion. The other and major part of the 
energy released during the first step is carried away 
unmeasured by the electron neutrino.  
The excited atomic hole states in $Dy$  decay into the 
ground state emitting either electrons or/and photons. A bolometric (calorimetric)  
measurement obtains the total energy of this second decay. 
The spectrum measured by the bolometer in the second step 
is given by the incoherent sum over the final hole states in Dy \cite{de}. 

\be
 \frac{d\Gamma}{dE_c} \propto (Q - E_c)\sqrt{(Q-E_c)^2 -m_{\nu}^2} \\
*\sum_{f=f'} \lambda_{0}B_f \frac{\Gamma_{f'}}{2\pi} 
\frac{1}{(E_c - E_{f'})^2 +\Gamma_{f'}^2/4} \label{decay}
\ee

The quantity $\lambda_{f0}$ contains the nuclear matrix element and is defined in 
eq. (\ref{trans0}). The total energy available for the decay, Q, is divided in the 
first step into the excitation energy $ E_c$ of $^{163}Dy$ 
and the energy of the emitted neutrino 
$E_{\nu} = \sqrt{m_{\nu}^2 \ + \ p_{\nu}^2} $. The excitation energy of Dy is then:

\be
 E_c \ = \  Q - E_{\nu} \ \  and \ \  
 E_c(max) =  \lim_{p_\nu \to 0} \ E_c\  = \ Q -m_\nu  \label{lim}
\ee

So the upper end of the decay spectrum in the calometric measurement is for 
all excitations of $^{163}Dy$ the same. 
The maximum available energy for the sum of 
the secondary processes (X-ray decay, Auger electron emission) measured by the bolometer 
is $E_c = Q -m_{\nu}$. 
This allows to determine the electron neutrino mass 
from electron capture due to a suppression  
 of counts around the interval $<\ Q\ -\ m_\nu\ , \ Q \ >$ \cite{Blaum, Gastaldo3}.
Since the finite neutrino mass is a very minor effect on the calorimetric 
spectrum, not only a very precise Q value must be determined, 
but one needs also, since the energy resolution is limited, 
the form of the spectrum at the upper end accurately. 
The purpose of the present work is to determine this spectrum and 
the effect of a finite neutrino mass on the spectrum, 
so that one has a chance to determine the 
mass of the electron neutrino. Among other effects we will show, 
that the bolometer spectrum depends 
sensitively on the overlap and exchange corrections. 
They determine the relative contributions
 of the excited hole states in $^{163}Dy$ on the spectrum. To study these effects 
is one of the purposes of this work. 

The sum f = f' in eq. (\ref{decay}) runs in principle over all 
occupied $s$ and $p_{1/2}$ electron levels in Holmium 
(see table \ref{psi}). Due to energy conservation the sum is constraint 
 to states $3s_{1/2}$ and $3p_{1/2}$ and higher. The sum runs also 
correlated over $f'$ the hole states in Dy with 
$ |f'>\ = \ |n, \  \ell,\ j> = |f>\ $ as a single sum over states  with the 
same quantum numbers. The reasons is that the 
captured electron in Ho must have for the leading 
overlap term the same quantum numbers as the hole in Dy. Contributions with $ f \neq f' $ 
are quadratic in the small "non-diagonal" 
overlaps (see table \ref{Dy3s}) and are neglected.   $ E_{f'}$ and 
$\Gamma_{f'}$ are the energies and the width of the hole states in $^{163}Dy$ 
given in eq. (\ref{BE}). The overlap 
and exchange correction $B_f$ is defined in eq. $(\ref{Bf})$ and includes a sum over the 
capture states in Holmium $i = 1s,\  2s,\  3s,\  4s,\ 5s,\  6s $ and $ 2p_{1/2}, \ 3p_{1/2}, \ 
4p_{1/2}, \ 5p_{1/2}$ (see table \ref{psi}).  
But as already mentioned due to energy conservation 
the sum for captured electrons in  $^{163}Ho$ is restricted to $3s_{1/2}$ 
and to $3p_{1/2}$ and higher.  All energy allowed occupied $s$ and $p_{1/2}$ 
levels in Ho  are included in this work. 

The effective anti-neutrino mass
$m_\beta$ measured in the Tritium decay  
and the neutrino mass to be measured 
in electron capture is given by
\begin{eqnarray}
m_\beta &=& \sqrt{\sum_{k=1}^3 |U_{ek}|^2 m^2_k} \nonumber\\
&\approx& m_\nu. \label{m}
\end{eqnarray}
The last approximate equal sign is obtained under the assumption 
 $m_\nu \approx m_1 \approx m_2 \approx m_3$.

The neutrinoless double beta decay \cite{GERDA, Simko, Fang} 
determines the effective Majorana neutrino mass:

\be
|< m_{\beta \beta}>| = | m_1 |U_{e1}|^2 + m_2 |U_{e2}|^2 \cdot e^{ i\alpha_{21}}
+ m_3|U_{e3}|^2 \cdot e^{i\alpha_{31}}| 
\ee

Here the angles $\alpha_{21} $ and $\alpha_{31}$ yield  the Majorana phases $e^{i\alpha} $.
For a time reversal symmetric theory they must be real and can have the values $ \pm 1 $ only. 
The coefficients $ U_{ej} $ of the Pontecorvo-Maki-Nakagawa-Sakata mixing matrix 
describe the transformation from mass $ \nu_1, \nu_2, \nu_3 $ 
to flavor eigenstates $ \nu_e, \nu_\nu, \nu_\tau $ of the neutrinos. 

\be
\left (\begin{array}{c} \nu_e\\ \nu_\mu \\ \nu_\tau  \end{array} \right)  = 
\left( \begin{array} {ccc} 
U_{e 1} & U_{e 2} & U_{e 3} \\
U_{\mu 1}& U_{\mu 2} & U_{\mu 3} \\
U_{\tau 1} & U_{tau 2} & U_{\tau 3} 
\end{array} \right)
\left (\begin{array}{c} \nu_1\\ \nu_2\\ \nu_3 \end{array} \right) \label{flavor}
\ee

\section{Description of Electron Capture}

The details of the evaluation of the electron capture probability 
are given by Bambynek et al. \cite{bam}.
$\lambda_{f0}$ is the electron capture probability assuming a 
complete identity and orthonormality between the 
electron states of the parent nucleus $(^{163}Holmium)$ and of 
the daughter nucleus $(^{163}Dysprosium)$. In our case:

\be
 \lambda_{0}  \propto G^2 \xi |\psi_{3s1/2}(R)|^2  \label{trans0}
\ee

G is the weak Fermi coupling constant, 
$ \xi $ the nuclear matrix element squared and $\psi_{3s1/2}(R)$ is the lowest 
energetically allowed single 
electron wave function of the parent at the 
nuclear radius (see table \ref{psi}) for the capture 
process of eq.  (\ref{reaction}). The radial part of the 
spinor wave function is defined and normalized as:

\begin{eqnarray}
 \psi_A(r) = \frac{1}{r} \left(\begin{array}{c}P_A \\ Q_A \end{array} \right) 
\ \ \ \ \ \ \ \ \ \  \hspace{3 cm} \nonumber \\
<A|B> = \int_{0,\infty} \Big( P_A(r) \cdot P_B(r) + Q_A(r) \cdot Q_B(r) \Big) \cdot dr 
= \delta_{A,B} \label{WF}
\end{eqnarray}

The overlap of wave functions with the quantum numbers $|A> = |n, \ell,j> $ 
in the parent atom and the quantum numbers $ |B'> = |n', \ell, j> $ in the daughter 
is given by (see table \ref{Dy3s}):

\begin{eqnarray}
<A|B'> = \int_{0,\infty} \Big( P_A(r) \cdot P_{B'}(r) + Q_A(r) \cdot Q_{B'}(r) \Big) \cdot dr 
=  overlap(A,B')       \label{Over}
\end{eqnarray}

Until now the probability of electron capture was  normally calculated 
with the captured electron wave function at the origin  
for $P_{ns}(r)$ and $ Q_{np1/2}(r) $. All the other electron wave 
functions disappear at  $ r \ = \ 0$.  In reality the wave functions have to be 
integrated in the nuclear matrix element over the nucleus. In almost all previous 
calculations these wave functions of the captured electrons are taken at the origin. 
In the present work as a better approximation the upper 
spinor amplitude for capture from the states $ns_{1/2}$  
and the lower amplitude for capture from the $np_{1/2}$ levels in Holmium are 
taken at the nuclear radius (\ref{Ra}). The finite size of the nucleus allows with 
reduced probabilities also capture from other than $ns_{1/2}$ and $np_{1/2}$ states 
(see table \ref{Vat}).

\be
R = 1.2 \cdot A^{1/3} \ [fm] = 2.2676\cdot 10^{-5} A^{1/3} \ [au] 
\rightarrow  1.2386\cdot 10^{-4} \ [au]\ \  for Ho;       \label{Ra}
\ee

We assume, that the total atomic wave function 
can be described by a single Slater determinant. 
$ B_f $ in eq. (\ref{decay}) takes into account the overlaps and the exchange 
terms between the parent $ |G> $ and the daughter atom in 
the state $ |A'_f> $ with a hole in the electron 
state $|f'>$. 

\be
 B_f = |\sum_i \psi_i(R) <A'_f|a_i|G>|^2/|\psi_{3s1/2}(R)|^2  \label{Bff}
\ee

Here $\psi_f(R) = P_{ns1/2}(R)/R$ or $Q_{np1/2}(R)/R $ are 
electron wave functions at the nuclear radius  in the parent 
state (\ref{WF}), which have the largest overlaps (\ref{Over})  
with the final electron hole states in the 
daughter nucleus. In eq. (\ref{Bff}) one has to divide out 
the wave function of the captured electron in Holmium 
 $|\psi_i(R)|^2 $ from eq.(\ref{trans0}) 
contained in eq. (\ref{decay}).
 The sum i in eq.$(\ref{Bff})$ runs over the occupied and energetically allowed  
 $ 3s,\ 4s,\ 5s,\ 6s,$ and $3p_{1/2}, \ 4p_{1/2}, \ 5p_{1/2}$,
 electrons states in $^{163}Ho$. Due to energy conservation 
 only captures of $3s_{1/2}$ and $3p_{1/2}$ states and higher are allowed. 

One assumes generally, that the initial $|G>$  and final $|A'_f> $ atomic 
wave functions can be given as Slater determinants. 
They are in this work calculated selfconsistently with 
a hole in state $i$ in eq. (\ref{Bff}) and 
with a single Slater Determinant Dirac-Fock code developed by Grant \cite{Grant} 
and modified by Desclaux \cite{Desclaux} and Ankudinov et al. \cite{Ankudinov}.

Configuration mixing affects mostly the valence electrons, while electron 
capture (EC) involves mainly the inner electrons. Thus a representation 
of wave functions by Slater determinants should be a good description for 
the EC process. But configuration mixing could lead to small satellite lines. 
The Slater wave functions of the initial Holmium in the
 ground state $ |G> $ and the excited electron hole states $ |A'_f> $ in 
Dysprosium read in second quantization:

\be
  |G> =  a_1^{\dagger} a_2^{\dagger} a_3^{\dagger}... a_Z^{\dagger} |0> \label{G}
\ee

\be
  |A'_f> =  a'^{\dagger}_1 a'^{\dagger}_2... a'^{\dagger}_{f-1}a'^{\dagger}_{f+1}... 
	a'^{\dagger}_{Z} |0> \label{A}
\ee

The primes' indicate the single electron spinor creation operators for the daughter nucleus 
(Dysprosium) with an electron hole in the single particle state $|f'> $. 
The following expression has to be calculated 
with the help of Wick's theorem $ \cite{Wick}$.  

\be
  \sum_i \psi_i(R) <A'_f|a_i|G> = \sum_i \psi_i(R) \cdot <0| a'_Z a'_{Z-1}...
	a'_{f+1}a'_{f-1}...a'_1 \cdot a_i \cdot 
	a_1^{\dagger} a_2^{\dagger} a_3^{\dagger}... a_Z^{\dagger} |0> \label{Wick}
\ee 

In eq.(\ref{Wick}) and also in later expressions for capture from Holmium $s_{1/2}$  
states the upper spinor components $P_{ns1/2}(R)/R$ and 
for capture from $p_{1/2}$ states  the 
lower spinor components $Q_{np1/2}(R)/R$  are taken 
for $\psi_i(R)$ at the nuclear radius (\ref{Ra}) 
in Holmium. The  
vacuum expectation value in eq. (\ref{Wick}) has now to be fully Wick contracted \cite{Wick}. 
Although the single electron 
wave functions are different for the parent (Holmium) and the daughter atom  
(Dysprosium) with a hole in $|f'>$, the corresponding 
electron states $|k>$ in the parent and $|k'> $ 
in the daughter with the same quantum numbers $|n,\ell, j>$  
have still the largest overlap (\ref{Over}) (see table \ref{Dy3s}). 
Contractions $<k'|m>$ with radial quantum numbers
 $ n_{m} $ and $ n'_{k'} $ different are small. The overlaps of parent and daughter 
states with different $ \ell$ and  $j$ are zero.  We neglect all terms with two or more 
such small single particle overlaps $<n', \ell, j|n, \ell,j>$. 
In case one obtains several large 
non-diagonal $ n \neq n'$ overlaps, such higher order terms must be 
included according to the vacuum expectation 
value (\ref{Wick}) with Wick's \cite{Wick} theorem.
 But the non-diagonal overlaps are normally about  two  or even three orders 
of magnitude smaller and quadratic terms in these small quantities can be neglected.
The maximum value of all the non-diagonal overlaps is $<3s'|4s>  =  0.023$. 
At this point one can already conclude, that the probability to excite 
two holes in the $^{163}Dy$ atom is less than $10^{-3}$ smaller, than exciting only one hole. 
A detailed discussion is given in the next section (See eq. (\ref{con}).)

\be
 F_{0f} = \psi_f(R)(-)^{f+1} \prod_{k \neq f}<k'|k> \label{F0f}
\ee

\begin{eqnarray}
 F_f = \sum_b \psi_b(R) <A'_f|a_b|G> \approx \sum_{b = 1...Z} \psi_b(R)  
  [(-)^{f+1} \delta_{b,f}  \prod_{k \neq f}<k'|k>  +                       \nonumber\\ 
	(-)^f \delta_{b \neq f} <b'|f> \prod_{k \neq (f,b)} <k'|k>]   \label{Ff}
\end{eqnarray}
\\
The index $b$ runs over the states $i$ from eq. (\ref{Wick}), 
from which the electrons are captured. 
The amplitudes $F_{0f}$ and $F_f$ are listed in table \ref{FNS}. 
If one wants to include the final 
excited daughter atom Holmium with two holes in $f_1$ and $f_2$ 
and an additional occupied electron state $c $, which can be a bound 
or a continuum state, one must add to the amplitude $F_f$ the amplitude ( \ref{con}). 

\begin{eqnarray}
 F_f(2\  holes) = \sum_b \psi_b(R) <A'_f(2\ holes)|a_b|G> 
\approx \sum_{b = 1...Z} \psi_b(R)  (-)^{f_2 - f_1 + Z}  \nonumber \\
 \cdot[<b'|f_1><c'|f_2>  -  <b'|f_2><c'|f_1>]  
	\cdot \prod_{k=1..Z \neq (f_1,f_2,b)}<k'|k>   \label{con}
\end{eqnarray}

Here again k and k' and also b and b' stand for the same quantum numbers in the parent and the 
daughter atom. The sum over b runs also over b'.  If $c$ is a 
continuum state (correctly normalized) one speaks of `shake off' into the continuum. 
Since now two "non-diagonal" overlaps are involved in eq. (\ref{con}) 
with $<b'|f_1> $ and $<c'|f_2>$ 
and the corresponding exchange term interchanging  $f_1$ and $f_2$ with a minus sign, 
this probability is by more than a factor $10^{-3}$ smaller 
as mentioned already above (for the overlap amplitudes see table \ref{Dy3s}) 
and is neglected. This expression  is here for 
the first time given analytically.  Eq. (\ref{Ff}) and (\ref{con}) 
describe the first step of the 
electron capture process, where the energy is 
carried away by the neutrino. The deexcitation of the 
hole or the two holes states in the second step can be measured 
using micro-calorimeters, as has been demonstrated by ECHo 
\cite{Gastaldo1, Gastaldo2, Blaum, Ra, Gastaldo5}. The hole 
states in $^{163}Dy$ deexcite due to X rays 
(smaller probabilities for the smaller transition energies in outer shells),
 Auger transitions (increasing  contributions  for outer shells)  
and Coster-Kronig transitions \cite{de}. 
The energy release in the first step by the neutrino escapes detection. 

Using eqs. (\ref{F0f}) and (\ref{Ff}) one can calculate the overlap 
and exchange corrections for eq. (\ref{decay}).

\be
 B_{0f} = |F_{0f}|^2/|\psi_{3s1/2}(R)|^2  \label{B0f}
\ee

\be
 B_{f} = |F_{f}|^2/|\psi_{3s1/2}(R)|^2  \label{Bf}
\ee

$B_{0f}$ is the correction including only the overlaps between 
the initial and the final atom, while
 $ B_{f}$ includes the overlap and exchange corrections. 
Bahcall $\cite{Bahcall1, Bahcall2} $ was the first to 
include the overlap and exchange corrections. 
Faessler et al. $ \cite{Fae1}$ included also selfconsistently the effect of holes in 
the daughter atom on the electron wave functions. 
\newline 
A first orientation is obtained, if one sets the `diagonal' 
overlaps $<k'|k> $ of electron orbitals with the same quantum 
numbers $n,\  \ell, \ j$ in the initial 
and the final atom equal to unity and the $'non-diagonal'$ 
overlaps equal to zero. In this approximation 
$|b>$ is the electron removed  from the parent (Holmium). 
But the state $|b'> $ in the overlap 
is an electron in the daughter (Dysprosium). $|f'>$ is  
the hole state in the daughter.  But for the overlap $<b'|f>$ in eq. (\ref{Approx}) 
the corresponding electron state in the parent is needed.
It seems, that Vatai \cite{Vatai,Vatai2} used the approximation putting 
all "diagonal" overlaps to unity $<k'|k> = 1.0$ and 
neglecting the exchange contributions. 
(Here labeled by "Approx0" and also some times 
called the Vatai approximation \cite{Vatai,Vatai2}, 
used also by De Rujula in ref. \cite{de}.)

\be
 F_{0f}(Approx0) = \psi_f(R)    \label{Approx0}
\ee

In the Vatai approximation \cite{Vatai,Vatai2}, used also 
by De Rujula \cite{de}, the overlap and exchange 
correction factor $B_f$ of eq. (\ref{decay}) simplifies.

\be
 B_{0f}(Approx0) = (\psi_f(R)/\psi_{3s1/2}(R))^2    \label{Over0}
\ee

The subscript $f$ indicated the atomic orbital 
of the parent, from which the electron is captured. 
Vatai and De Rujula \cite{de} use this expression at the origin $R\ = \ 0.0$. 
In a slightly improved version one can also include exchange terms.

\begin{eqnarray}
 F_f (Approx) = \psi_{f}(R)  - \sum_{b \neq f} \psi_b(R) <b'|f>   \label{Approx}\\
 B_f(Approx) = (F_f(approx)/\psi_{3s1/2}(R))^2
\end{eqnarray}

The overlaps listed in table $ \ref{Dy3s} $ show , that 
indeed this approximation can give a first quite reasonable approximation. 
But for all results reported here, if not otherwise stated, 
we use the exact expressions. The relative plus and minus signs of states, which mix
for different $n$ but the same $\ell$ and $j$, are essential. 
But  an overall minus sign is irrelevant.

The single electron states occupied in the Holmium ground state are:

\begin{eqnarray}
 (|1s_{1/2}>)^2,  (|2s_{1/2}>)^2,  (|2p_{1/2}>)^2, \nonumber\\ 
 (|2p_{3/2}>)^4, (|3s_{1/2}>)^2, (|3p_{1/2}>)^2,  \nonumber\\
 (|3p_{3/2}>)^4,(|3d_{3/2}>)^4,(|3d_{5/2}>)^6, \nonumber\\
   (|4s_{1/2}>)^2,(|4p_{1/2}>)^2, (|4p_{3/2}>)^4,  \nonumber\\
 (|4d_{3/2}>)^4,(|4d_{5/2}>)^6 ,   (4f_{5/2}>)^6, \nonumber\\ 
 (|4f_{7/2}>)^5, (|5s_{1/2}>)^2,(|5p_{1/2}>)^2,           \nonumber\\
  (|5p_{3/2}>)^4 ,(|6s_{1/2}>)^2; \label{Hog}
\end{eqnarray}

Using like Vatai \cite{Vatai2} -  but in our case in a relativistic treatment -  
the approximation as in eq. (\ref{Approx0}) one get a simple expression for
$F_f(approx0)$ used also by De Rujula \cite{de}. The amplitudes $F_f$ and 
its approximations as described above are given in table \ref{FNS}. \\

The 163 Dysprosium ground state with $Z = 66 $ has only 4 
electrons in $ (|4f_{7/2}>)^4$. The small Q-value 
of electron capture in 163 Holmium to 163 Dysprosium 
of \  $Q \ = \ 2.3 \ to \ 2.8 \ keV$ is optimal 
for the determination of the electron neutrino mass. 
It also restricts the excited hole states in Dysprosium 
to $3s_{1/2}\  (M_1)$ and higher.
The experimental excitation energies of the hole states and there 
widths are given in eq.(\ref{BE}) taken from reference 
\cite{Gastaldo3}. (See also \cite{Wang,Ra, Audi}.)

\begin{eqnarray}
E(3s_{1/2}, M_1) = 2.040 keV; \  \Gamma = 13.7 eV;  \nonumber \\
E(3p_{1/2}, M_2) = \  1.836  keV; \  \Gamma =  \  7.2 eV;  \nonumber \\
E(4s_{1/2}, N_1) = 0.411 keV; \  \Gamma =  \ 5.3 eV;  \nonumber \\
E(4p_{1/2}, N_2) = 0.333 keV; \  \Gamma = \  8.0 eV;  \nonumber \\
E(5s_{1/2}, O_1) = 0.048 keV; \  \Gamma =  \ 4.3 eV;   \label{BE}
\end{eqnarray}

For the excited Dy atom with  a hole in a singly occupied 
$ (|3s_{1/2}>)^1$ the number of electrons in $ |4f_{7/2}>$ is five.  
As an example lets write down the expression (\ref{Ff}) for capture into the final
$ M1 \equiv 3s$ state .
\begin{eqnarray}
 F_{3s,Dy} = \psi_{3s,Ho}(R)<1s'|1s>^2 \cdot <2s'|2s>^2 \cdot <3s'|3s>^1 \nonumber\\\
 \cdot <4s'|4s>^2 \cdot <5s'|5s>^2 \cdot <6s'|6s>^2       \nonumber\\
\cdot <2p'_{1/2}|2p_{1/2}>^2 \cdot <3p'_{1/2}|3p_{1/2}>^2 \nonumber\\ 
\cdot <4p'_{1/2}|4p_{1/2}>^2 \cdot <5p'_{1/2}|5p_{1/2}>^2 \nonumber\\
\cdot <2p'_{3/2}|2p_{3/2}>^4 \cdot <3p'_{3/2}|3p_{3/2}>^4 \nonumber\\
\cdot <4p'_{3/2}|4p_{3/2}>^4 \cdot <5p'_{3/2}|5p_{3/2}>^4  \nonumber\\
 \cdot <3d'_{3/2}|3d_{3/2}>^4 \cdot <4d'_{3/2}|4d_{3/2}>^4 \nonumber\\
 \cdot <3d'_{5/2}|3d_{5/2}>^6 \cdot <4d'_{5/2}|4d_{5/2}>^6 \nonumber\\ 
 \cdot <4f'_{5/2}|4f_{5/2}>^6 \cdot <4f'_{7/2}|4f_{7/2}>^5  \nonumber\\ 
 -  \sum_{b=1...Z \neq 3s} \psi_{b,Ho}(R)<b'|3s>\prod_{k \neq (3s,b)} <k'|k>   \label{F3s}
\end{eqnarray}

The primed $|k'>$ states are in the daughter (Dysprosium)  atom and the unprimed single 
electron orbits in the parent atom (Holmium) $|k> $.
These two orbits with the same $ n, \ \ell \  and\ j$ 
have still a large overlap close to unity (see table \ref{Dy3s}).  
The sum over b runs also over b' and is restricted 
by orthogonality to the $ns_{1/2}$ electrons in Ho excluding the $3s_{1/2}$ level.  
$|f'>$ are the empty electrons states in the daughter 
atom (Dysprosium).
The wave functions $\psi_b(R) = \psi_{b,Ho}(R)$ 
have to be taken at the position of the nucleus in the Holmium atom. More exactly they should 
be integrated over the nucleus with the weak interaction Hamiltonian with 
the weight $r^2$. Since the surface of the nucleus has the 
largest weight for EC the values of $ \psi_{b,Ho}(R)$ are taken in this work for the upper 
spinor component for the $ ns_{1/2},\  P_{s}(R)/R$ and for the lower 
component $np_{1/2}, \ Q_{p1/2}(R)/R$ at the nuclear radius R.

In the selfconsistent Dirac-Fock approach \cite{Grant, Desclaux, Ankudinov}
we include the finite size of the nucleus 
as Fermi distribution with a diffuseness adjusted to electron-nucleus scattering data
 \cite{Frauen}.

\be
 \rho(r) = \frac{\rho_0}{1 + exp[ (r - R)/a]}  \label{dis}
 \ee
 \be
 a = 0.546 \  [fm] = 1.0318 \cdot 10^{-5}\ [au];  \label {dif}
\ee

The length for the atomic units $ [au]$  is the Bohr radius of Hydrogen.
\be
  1 [au] = 0.529177\cdot10^{-8}\ [cm];  \label {au}
\ee

The charge parameter of the nucleus $ \rho_0$ is normalized in Holmium to a total charge 
of $Z= 67$ and in Dysprosium to $Z = 66$ protons in the nucleus.
The diffuseness $a$  is taken from the book of Fraufelder and Henley 
 \cite{Frauen}.
 The wave function of the 
captured electron in the parent atom (here Holmium with $ Z = 67$ \ and \ $A= 163$) 
should be integrated over the nucleus. In almost all  previous 
calculations  of the overlap and exchange contributions 
for electron capture either a point nucleus is assumed or/and 
the captured electron wave function is taken at the origin. 
In this work the nucleus is treated with a finite charge distribution (\ref{dis}) and  the 
value of the captured electron wave function is taken at the nuclear radius. 
For capture from $s_{1/2}$ states the upper $P_{ns1/2}(R)/R$ and 
for capture from $p_{1/2}$ states the lower radial spinor 
waves $Q_{np1/2}(R)/R$ (\ref{WF}) are used. An overall sign in eq. (\ref{Ff}) 
and in eq. (\ref{F3s}) is accidental and is also irrelevant, 
since the expression enters as absolute squared. 

Bahcall \cite{Bahcall1,Bahcall2} studied non-relativistically 
in lighter atoms $Z = 14  $ to $37$ 
the overlap and exchange corrections. The non-relativistic 
treatment restricts to capture of $ns_{1/2} $ electrons. 
In his case the  wave functions of the captured electrons are taken at the origin 
$ \psi_{ns1/2}(0)$ and a point nucleus is assumed. 
Hole states for the determination of the electron 
wave functions of the daughter atom and also the multiplicity of 
several electrons in the same orbit are not included for the overlap and exchange corrections.  
Faessler et al. \cite{Fae1} were the first to include hole states for the selfconsistent
 determination of the electron wave functions  in the daughter nucleus. 

The approximate expressions used by Bahcall \cite{Bahcall1, Bahcall2} 
for much lighter systems 
in a non-relativistic treatment  are:

\begin{eqnarray}
 F_{1s'} (Bahcall) = <2s'|2s><3s'|3s>\psi_{1s1/2}(0) \hspace{2 cm} \nonumber \\
-<2s'|1s><3s'|3s>\psi_{2s1/2}(0) 
- <3s'|1s><2s'|2s>\psi_{3s1/2}(0)  \label{Bah1}
\end{eqnarray}

\begin{eqnarray}
 F_{2s'} (Bahcall) = <1s'|1s><3s'|3s>\psi_{2s1/2}(0) \hspace{2 cm} \nonumber \\
-<1s'|2s><3s'|3s>\psi_{1s1/2}(0) 
- <3s'|2s><1s'|1s>\psi_{3s1/2}(0)  \label{Bah2}
\end{eqnarray}

Vatai \cite{Vatai, Vatai2} derived using explicitly 
Slater determinants the non-relativistic equivalent 
of the formulation given here. We used second quantization. 
He includes for the overlaps (\ref{Over}) in his theoretical formulation 
the multiplicity of electrons in the same orbit, but 
as it seems not in the numerical calculations, 
where he did set all overlaps of equivalent orbitals $<k'|k> \equiv 1.0 $ of the parent $|k>$ 
and the daughter $|k'>$ equal to unity, so that 
the multiplicity problem is not relevant for him. (See eqs. (5) and (6) of ref. \cite{Vatai2}). 
The  existence of holes in the final atom seems not to be included  
in the numerical treatment, although the problem is discussed. 
(See ref. \cite{Vatai} page 542 (ii)). 
The nucleus seems be treated as point like. (See  \cite{Vatai} page 545 chapter 3.1.)  

Lets assume one has measured or calculated  the electron capture probability 
$P(n_{0},\ \ell_{0},\ j_0)$ into the daughter (Dysprosium) state $ |n_{0}, \ell_{0},\ j_0> $, 
what is then the probability  
for EC into an other final hole state $P(n,\ \ell,\ j)$?

\be  
 Br(n,\ \ell,\  j/ n_0, \ \ell_0,\ j_0) = \frac{P(f = n,\ \ell,\ j)}
{P(f_{0} = n_{0},\ \ell_{0},\ j_0)} 
 =  \frac{B_{f=n\ \ell \ j} \cdot |\psi_{f}(R)|^2}
{B_{f_{0} = n_{0}\ \ell_{0}\ j_0} \cdot |\psi_{f0}(R)|^2}                  \label{rel}
\ee
with $ B_f $ from eqs. (\ref{Bff}), (\ref{Ff}) and (\ref{Bf}). 
These values are listed in table \ref{Rel}.
\vspace{1 cm}

\section{Numerical Results.}

The theoretical calorimetric spectrum of the 
bolometer for the decay of the single hole states in Dysprosium 
is shown in figure \ref{Spectrum1} with the relative overlap and exchange 
corrections $Br(n, \ \ell, \ j/ \ n_0, \ \ell_0, j_0)$ eq. (\ref{rel}) 
given in  table \ref{Rel}.  
By choosing all overlap and exchange correction 
factors unrealistically in eq. (\ref{decay}) $B_f\  =\  Br\  = \ 1.0$ 
the corresponding results a displayed in figure \ref{Spectrum2}.  
With this choice for the factors $B_f\ = \ 1.0$ they  
disappear from eq. (\ref{decay}). Loosely speaking one can say that this choice 
yields results "without overlap and exchange corrections". Figure \ref{Spectrum2}
 and \ref{Spectrum279-280} show such results with this 
unrealistic choice of all $B_f \ = \ 1.0$ 
as an information for the reader. 
Figure \ref{Loredana} shows the experimental spectrum 
according to \cite{Gastaldo4, Gastaldo5}. 
A comparison of these three figures shows the 
need for the "overlap and exchange correction factors" $B_f$.  
\newline
To see also the spectrum in between the resonances, which stand out in figures 
\ref{Spectrum1} and \ref{Spectrum2} and also in 
the experimental spectrum fig. \ref{Loredana}, the 
logarithmic theoretical spectrum for the value $Q\  = \ 2.80 \ keV $ is shown 
in figure  \ref{Spectrum0-28log} for the neutrino mass $m_\nu\  = \ 0 \  eV$ 
with overlap and exchange corrections. 

De Rujula \cite{de} shows also the logarithmic spectrum 
for the Q-value $Q \ = \ 2.5 \ keV $ in his 
figure 12 with older values for the excitation energies of the hole states and their width 
in Dysprosium. Newer values \cite{Gastaldo3} used here are listed in eq. (\ref{BE}). 
To estimate the overlap corrections De Rujula used the "Vatai"-approximation,
called in the present  paper "Approx0" in eqs. (\ref{B0f}) and (\ref{Approx0})   
\cite{Vatai} and \cite{Vatai2}, which assumes 
a 100 percent overlap between the corresponding electron wave functions in Holmium 
and in Dysprosium and neglects all exchange corrections. 
He takes the electron wave functions  
in Ho at the origin from tables of reference \cite{Band}. The correction applied 
by De Rujula \cite{de} is then simply

\be  
Br(De \ Rujula; f/3s_{1/2})\ = \ (\psi_f(0)/\psi_{3s1/2}(0))^2  \label{RU}
\ee

 with $\psi_{3s1/2}(0) \ = \ \psi_{M1}(0)$ instead of the more 
exact expression of eq. (\ref{rel}) with (\ref{Ff}) and (\ref{Bf}). De Rujula 
takes the wave functions $\psi_f(0)$ and
$\psi_{3s1/2}(0)$  
at the origin in Holmium for the 
corresponding quantum numbers $|f>\ =\ | n,\ \ell, \ j>$ of the hole states in Dysprosium 
from ref. \cite{Band}.
In this work the wave functions are taken at the nuclear radius, 
since in the nuclear matrix element the electron wave functions are integrated over 
the whole nucleus with a weight factor $r^2$. The factor $r^2$ 
suppresses the influence of $\psi_f(r)$
at $ r\ = \ 0$. 

Our spectrum at the 
upper end  $<2.79 \ keV,\  2.80 \ keV> $ is given for the neutrino masses 
$m_\nu \ = \ 0,\ 2,\  5\ eV$ with and without the 
relative overlap and exchange corrections $Br$ 
defined in eq. (\ref{rel}) with (\ref{Bf}) (numerical values in table \ref{Rel} )
in figures \ref{Spectrum279-280B} and \ref{Spectrum279-280}. 
The end point of the theoretical calorimetric 
spectrum is determined by the Q value and the neutrino mass and thus is 
the same with and without overlap and exchange corrections. 
But the slope of the spectrum near the end point is slightly  
different. Due to the experimental 
finite energy resolution one cant determine the neutrino mass just 
by looking to the disappearance of the spectrum. One will need 
to perform a least square fit to the data of  the theoretical spectrum 
near the end point varying the Q-value,   
the neutrino mass and probably also the finite energy 
resolution including also a background. 
So figure 5 could be the starting point of such a fit. 

Table \ref{Rel} gives the probabilities for the $4s_{1/2}, 5s_{1/2}, 
3p_{1/2}, 4p_{1/2}$, and   $5p_{1/2} $  holes relative to the 3s hole in Dy  
with a vacancy in these states.  Due to the small $Q\approx 2.5 [keV]$ Q-value 
only $3s, 4s, 5s, 6s$ and $3p_{1/2}, 4p_{1/2}, 5p_{1/2}$ holes can be excited. 

In table \ref{Vat} the relative probabilities also for the excitation of 
other hole states in $^{163}Dy$ apart of $s_{1/2}$ and $p_{1/2}$ are given. 
These other hole states can be excited due to the description of the nucleus 
with a finite Fermi charge distribution (\ref{dis}), (\ref{dif}) and (\ref{au}).
Electron capture is proportional to the probability to find the electron inside 
the nucleus weighted with $r^2$. This emphasizes the role of the electron 
wave functions at the nuclear radius. There also 
other than $s_{1/2}$ and $p_{1/2}$ electron wave functions 
are different from zero.  For the determination of the 
relative probabilities in table \ref{Vat} the Vatai 
approximation (\ref{RU}) is used, since this approximation already shows, that 
capture from other than $s_{1/2}$ and $p_{1/2}$ states can be neglected.  
The results in table \ref{Rel} are calculated 
with the exact expressions (\ref{rel}).

\section{Neutrino Mass, Energy Resolution and the \\ 
Electron Capture Spectrum of $^{163}$Holmium.}

The experimental bolometer spectrum of EC in $^{163}Ho$ has a 
finite energy resolution characterized by the "Full Width Half Maximum (FWHM)".
This complicates the extraction of the neutrino 
mass from the upper end of the bolometer spectrum near the 
Q value. In figures \ref{Word19} and \ref{Word18} 
the effect of finite energy resolution is shown. 

Figure \ref{Word19} displays for neutrino masses 0 eV and 1 eV  the bare 
theoretical spectra and the two spectra  width FWHM (Full Width Half Maximum) =
 1 eV. After folding the complete $< 0.000 \ [keV]; 2.802\ [keV]>$ 
bare theoretical spectrum with a Gaussian the integral over the bare and 
the folded spectrum has the same value. But this is not the case for a 
finite folding interval,  here $<2.7500 \ [keV];\  2.8020\  [keV]> $. 
Strength from the left side of the lower bound of the  energy of the 
integral can not be moved into the folding area and strength from the 
folding interval is moved on the left side outside the integral below 2.75 keV 
and lost. The effect is for 1 eV FWHM a reduction of the folded spectrum practically 
independent of the assumed neutrino mass by $1\ \%$. A renormalization of the 
folded spectrum by multiplying it by a factor 1.01 can practically not 
been seen in the figure and thus is not shown. In addition such a 
renormalization factor should depend on the energy.  

But if one integrates the spectrum folded in 
the interval $<2.7500\ [keV];2.8020 \ [keV]>$
over a smaller  interval $<2.7500 \  + \ N*FWHM keV;\  2.820\  keV>$ 
with $N\ >\  5$ 
the integral is even slightly larger than the 
one over the bare spectrum. 
(Hardly visible in the figure and thus not included in it.)  
This is due to the rapid decrease of the 
spectrum with increasing energy at the upper end near the Q value. The folding moves 
more relative intensity to the right than to the left over the initial 
energy of the integration. Figure \ref{Word19} 
displays the last 80 mesh points for the energy interval $<2.7947 \ [keV]; 2.8020\ [keV]>$
corresponding to $7.3\  eV$ for a spectrum folded 
over the interval $<2.7500\ [keV]; \ 2.8020 \ [keV]>$. 

Figure \ref{Word18} shows the same as figure \ref{Word19} but for the $FWHM\ 
 =\  3\  eV$ for a Gaussian folded into the spectrum 
in the interval $<2.7500 \ [keV];\  2.8020 \ [keV]>$. 
The integral over the folded spectrum is now $3\ \%$ 
smaller than the same integral over the bare spectrum due to 
the three times larger FWHM of the Gaussian. But it is still so small, that a renormalization 
of the folded spectrum by a factor $ 1.03$ can hardly be seen in figure \ref{Word18} 
and thus is not shown here. 

To determine the neutrino mass from EC in $^{163}$Ho the  electron capture data 
must probably be adjusted with a maximum likelihood method 
varying  the neutrino mass, the Q-value and probably also the energy resolution 
to simulations like in figures \ref{Word19} and \ref{Word18}, 
since not only the neutrino mass, but also the Q value and the 
energy resolution are not known accurately enough. 
The overlap and exchange corrections affect 
the slope of the spectrum just below the Q value, thus these corrections must be included 
in such an analysis. 

\vspace{1cm}

\section{Technical Details.}

Figures \ref{Radius1} and \ref{Radius2} show the connection between the 251 
logarithmic arranged radial mesh points t and the radius in atomic units 
$[au \ = \ a_0\  =\  Bohr\  radius = 
0.529177\cdot10^{-8} [cm] ]$. Figure \ref{Holmium}  shows the upper amplitudes 
$P_{ns1/2}$ (\ref{WF}) for $n\  = \ 1,\ 2,\ ...6$ 
in Holmium as a function of the radial parameter t eqs. (\ref{rr0}),  
(\ref{rr1}) and (\ref{rr2}).

\be  
 t = \ln(r/r_0)/h;  \ \  h = 0.05; \ \ r_0 = 7.1469\cdot10^{-5} [au]. 
\ \ t \ = \ 1,2,...251  \label{rr0}
\ee  

Figure \ref{smallr} demonstrates,
that  at small radial distances from the origin the lower component $Q(r)$ for $2p_{1/2}$ 
is larger than the upper component P(r). 
Figure \ref{small} shows the amplitudes $P(r)/r$ and $Q(r)/r$ 
(\ref{WF}) for $1s_{1/2}$ and $2p_{1/2}$ 
and hints , that the lower component $Q_{2p1/2}(r)/r$ 
approaches a value different from zero at the origin. 

Table \ref{psi} shows the relativistic wave functions $P_{ns}(R)/R$ 
and $Q_{ns}(R)/R$ and also $P_{np1/2}(R)/R$ 
and $Q_{np1/2}(R)/R$ at the nuclear radius calculated and used in this work. These 
values are compared with the non-relativistic 
Froese-Fischer (FF) \cite{Froese} and the relativistic results of 
Mann and Waber (MW) \cite{Mann}  at the origin for the upper (u) and 
the lower ($\ell$) amplitude (\ref{WF}). At $r= 0$ 
the lower $ns_{1/2}$ and the upper $np_{1/2}$ components (\ref{WF}) are zero. 

Table \ref{Dy3s} lists the overlaps  $<n', s'_{1/2}|n, s_{1/2}>$ 
and  $<n', p'_{1/2}|n, p_{1/2}>$ of  $A=163$
Dysprosium states $<k'|$ with a hole in $3s'\ ( M'_1)$ with states $|k>$ 
in the parent Holmium. Table \ref{FNS} 
gives the amplitudes (\ref{Approx0}), (\ref{Approx}), (\ref{F0f}) 
and (\ref{Ff}) calculated with our relativistic values $\psi(R) $ and overlaps 
and also the corresponding results with the relativistic wave functions $\psi(0)$ tabulated 
by Mann and Waber \cite{Mann} at the origin and our overlaps 
(see for example  table \ref{Dy3s} ).

To calculate the electron capture correction factors $B_f$ (\ref{decay})  
\cite{Bahcall1, Bahcall2, Fae1, Vatai, Vatai2} for $^{163}_{67}Ho \rightarrow ^{163}_{66}Dy $ 
in eqs. (\ref{decay}), (\ref{Bff}) and (\ref{Bf}) one assumes, that the two ground states 
and the states of Dy with electron vacancies 
in the different states $ 1s_{1/2}\ (K),\ 2s_{1/2}\ (L_1),\ 3s_{1/2}\ (M_1),
\ 4s_{1/2} (N_1) \ and\  5s_{1/2}\ (O_1) \  $ and also 
$\ 2p_{1/2}\ (L_2),\ 3p_{1/2}\ (M_2),\ 4p_{1/2}\ (N_2) \ and\  5p_{1/2}\ (O_2)\  $ 
can be described by a selfconsistent 
Slater determinant. We use here selfconsistent relativistic 
wave functions for the ground states in Holmium and in Dysprosium and 
allow also explicitly electron 
vacancies in the specific final hole states using the Dirac-Fock code of Grant \cite{Grant} 
with modifications and simplifications by Desclaux \cite{Desclaux} 
and Ankudinov et al. \cite{Ankudinov}.
Relativistic effects contract the inner electron shells 
and increase in Z = 67 nuclei the amplitudes  
of the electrons at the nucleus $ \psi_{b,Ho}(0) $ by 
about a factor 2. In addition apart of the
 $\ell = 0 $ states also the $p_{1/2}$ states have, due to the lower  ( so called `small') 
amplitude, a finite probability to be at the nucleus. 
The correction factor $B_{n\ell\ j, Dy} $ depends only 
on the relative size of $ \psi_{n\ell \ j,Dy}(R)/
\psi_{n0,\ell0,\ j0,Dy}(R) $. The $ \psi_{n \ p1/2}(R)$ 
amplitudes of  $n\ p1/2 $ are $ 0.20$ to $0.25$ 
of the corresponding \  $ |n,\  s_{1/2}>\ $ states or 4 to 5\% 
for the probability relative to the s states. 
But care has to be taken, since in the interference terms the corrections are 
proportional to the amplitudes. In Holmium and Dysprosium a 
relativistic treatment is definitely needed. 
(See also reference \cite{Mukoyama}.)

For the ground state of Holmium the electron multiplicities 
the occupation of the different orbitals are given in eq. (\ref{Hog}). 
The occupation of the ground state and the excited hole states 
of Dysprosium are discussed after eq. (\ref{Hog}). 
The occupation of the in the ground states of $ 4f_{7/2} $ is $5/8$ in Ho and $4/8$ in Dy. 

The scale t is logarithmic  in the radial distance. 
The relation between the radius and the distance parameter t is 
shown in figures  \ref{Radius1} and \ref{Radius2}. The choice of t in  figures \ref{Holmium}, 
\ref{smallr} and \ref{small} displays more in detail   
the area around the nucleus, where the potential is changing fast. 
The 251 mesh points are defined as $t = 1,2,....251$ \cite{Ankudinov}:

\be  
 r = r_0 e^{ht} \ [au]\ ; \ \  t\ =\  1, \ 2, ...\ 251 \hspace{2 cm}                                      \label{rr1}
\ee

\be  
 r_0 = 7.14693\cdot10^{-5}\ [au]; \ \ ;  h = 0.05  \ \    for \ \ Holmium                               \label{rr2}
\ee

\be  
 r_0 = 14.33817\cdot10^{-5}\ [au]; \ \ ;  h = 0.05  \ \    for \ \ Dysprosium                              \label{rr3}
\ee
\vspace{1cm}

 To guarantee the accuracy needed  
\cite{Ankudinov}  we use double precision. 
To obtain the wave functions at the same meshpoints for Ho 
and for Dy we use fourth order Lagrange interpolation. 
The integrations are performed with Simpson. The logarithmic 
radial scale $\ t\  = \ 1, 2,...251$ \  for the figures is 
always translated to the convention for Holmium (\ref{rr2}).

\vspace{1 cm}

\section{Conclusion}
 
\begin{figure}[tp]
\begin{center}
\begin{minipage}[t]{17cm}
\epsfig{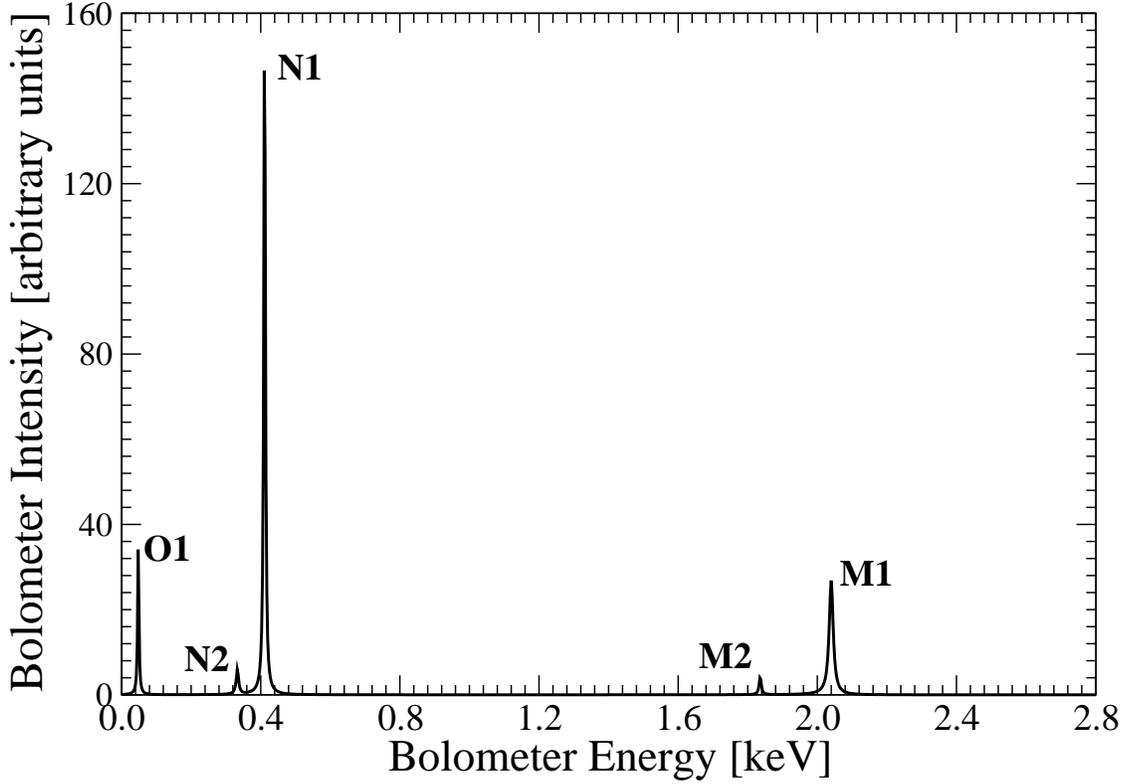}
\end{minipage}
\begin{minipage}[t]{16.5 cm}
\caption{The theoretical bolometer (calorimetric) spectrum of this work 
of the deexitation of the hole states in 
Dysprosium in the second step after the initial electron capture in Holmium. 
The deexcitation energy in the first 
step is carried away by the neutrino and is not measured.  
The bolometer sums up all energies emitted in the second step: X-rays, Auger electrons 
and the recoil of the Holmium nucleus. Since the recoil is in the order of meV, 
it can be neglected.  
The spectrum is based on eq. (\ref{decay}) assuming an incoherent 
deexcitation of the different hole states using experimental energies 
and width (\ref{BE}). The overlap and 
exchange correction $B_f$ is included according to 
to column 5 $Br$  of table \ref{Rel}. The Q value 2.8 keV 
and the hole binding energies and width 
eq. (\ref{BE}) are taken from \cite{Gastaldo3}. 
This spectrum compares well with the data from figure \ref{Loredana}. 
In the mean time the O1 resonance line has also been measured \cite{Gastaldo3}in good agreement 
with this theoretical results .
But since these experimental data are not yet published, 
it is not shown in figure \ref{Loredana}.
\label{Spectrum1}}
\end{minipage}
\end{center}
\end{figure}

\begin{figure}[tp]
\begin{center}
\begin{minipage}[t]{17cm}
\epsfig{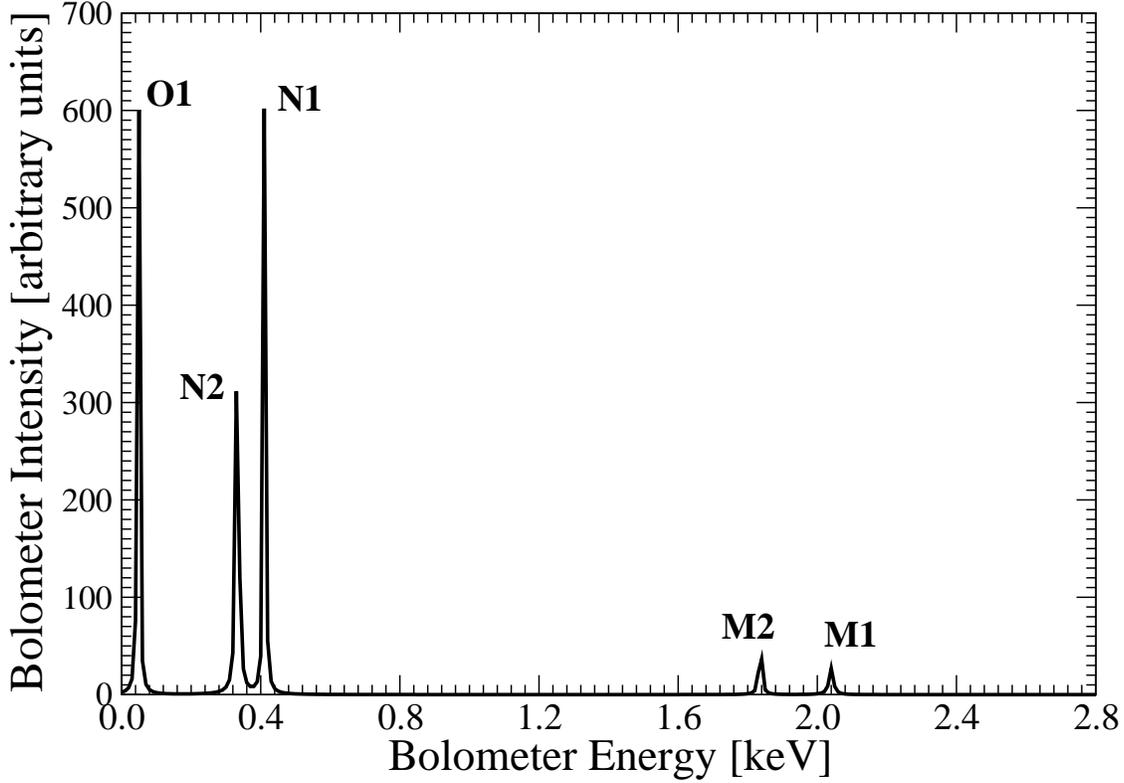}
\end{minipage}
\begin{minipage}[t]{16.5 cm}
\caption{Theoretical spectrum as in figure \ref{Spectrum1} 
but with the choice  $B_f = B_r = 1.0$ for the capture from all 
energetically allowed electron levels 
in $163 \ Holmium$. By this unrealistic choice the overlap and exchange coefficients 
do not show up anymore in eq. (\ref{decay}). Thus one could loosely speaking say
the results are calculated "without overlap 
and exchange correction" $B_f$. This unrealistic choice of $B_f$ 
"without overlap and exhange corrections" 
cant reproduce the data from figure \ref{Loredana}.
\label{Spectrum2}}
\end{minipage}
\end{center}
\end{figure}

\begin{figure}[tp]
\begin{center}
\begin{minipage}[t]{17cm}
\epsfig{file=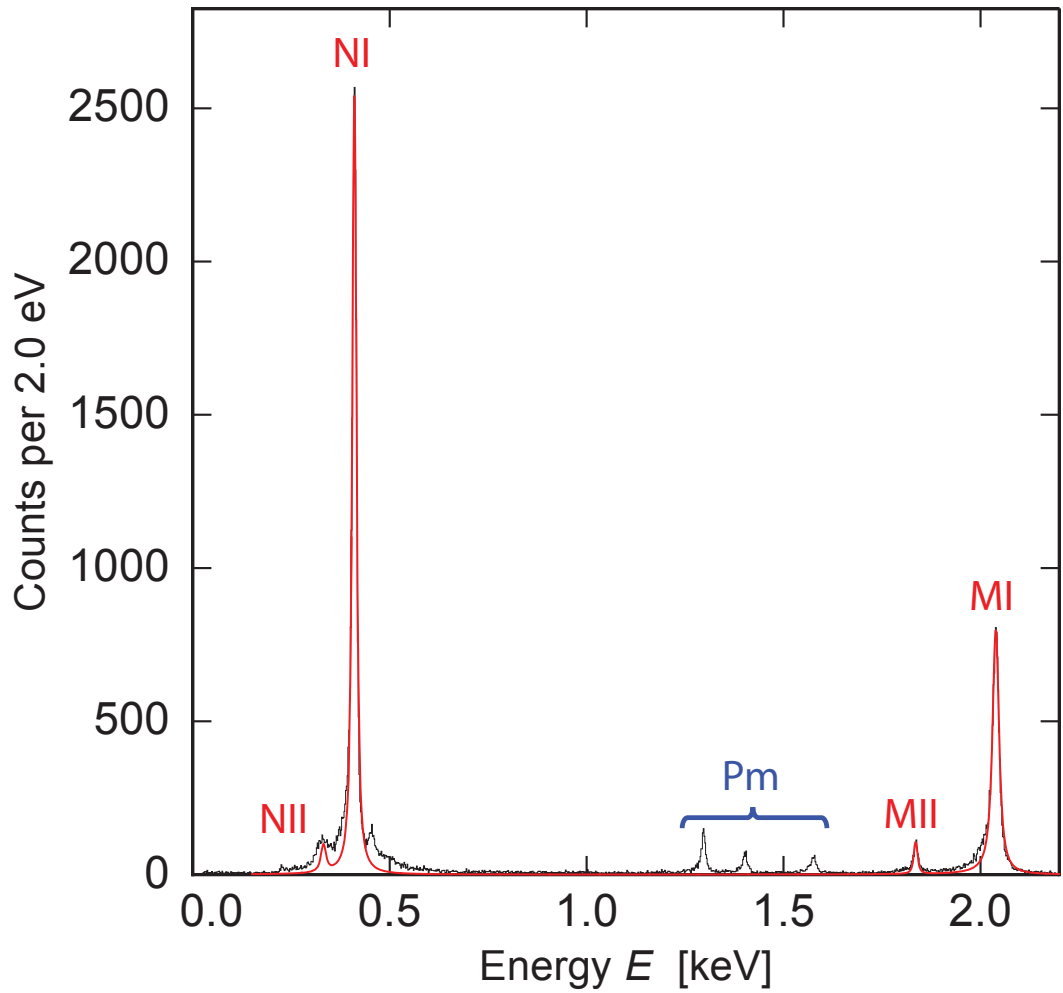,scale=0.9}
\end{minipage}
\begin{minipage}[t]{16.5 cm}
\caption{Experimental bolometer (calorimetric) spectrum of the deexitation of the hole states in 
Dysprosium in the second step after the initial electron capture in 
Holmium according to \cite{Gastaldo4}. 
\label{Loredana}}
\end{minipage}
\end{center}
\end{figure}

\begin{figure}[tp]
\begin{center}
\begin{minipage}[t]{17cm}
\epsfig{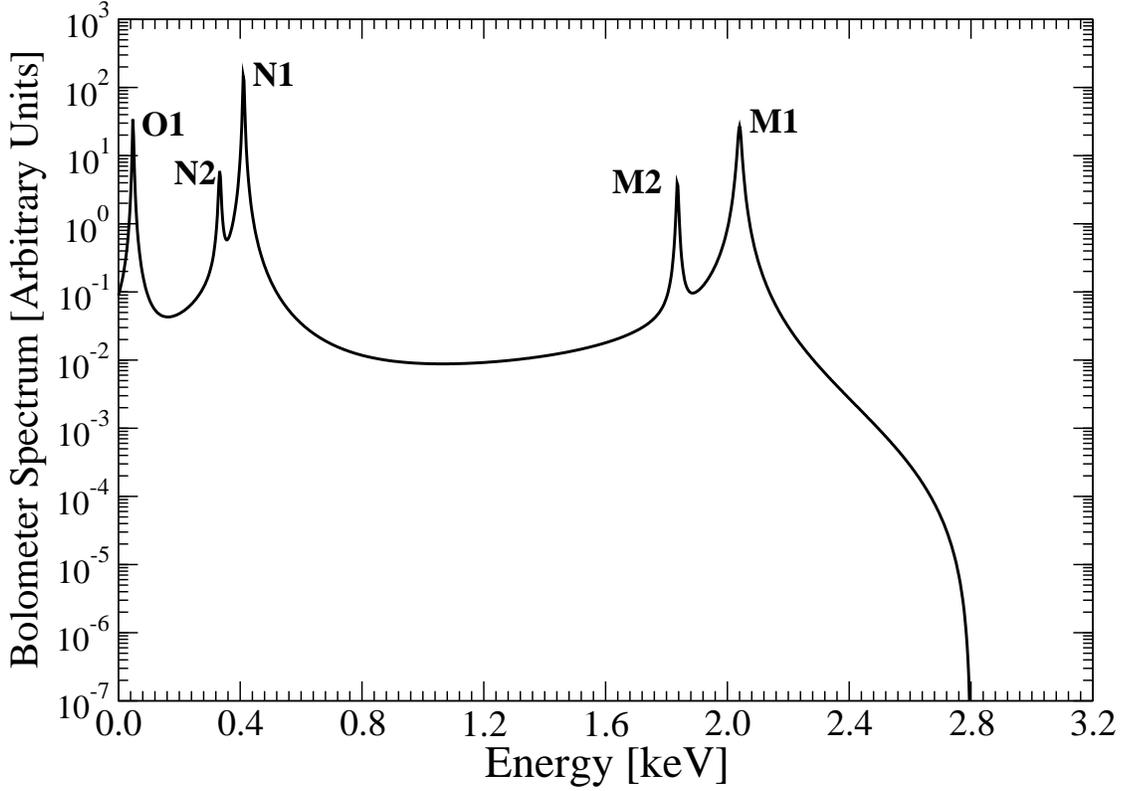}
\end{minipage}
\begin{minipage}[t]{16.5 cm}
\caption{Theoretical logarithmic bolometer (calorimetric) spectrum 
of the deexitation of the hole states in 
Dysprosium in the second step after the initial electron capture in Holmium 
for the Q value $2.8\  keV $ 
and the neutrino mass $ m_\nu \ = \ 0.0 \ eV$. The
overlap and exchange corrections $B_f$ of 
eq. (\ref{decay}) and eq. (\ref{Bf}) from table \ref{Rel} 
are included. At this scale of the energy 
resolution it is not possible to show the theoretical effect of a 
finite neutrino mass of the order of 1 eV.  For the effect of a finite 
neutrino mass see figure \ref{Spectrum279-280B}. 
\label{Spectrum0-28log}}
\end{minipage}
\end{center}
\end{figure}

\begin{figure}[tp] 
\begin{center}
\begin{minipage}[t]{17cm}
\epsfig{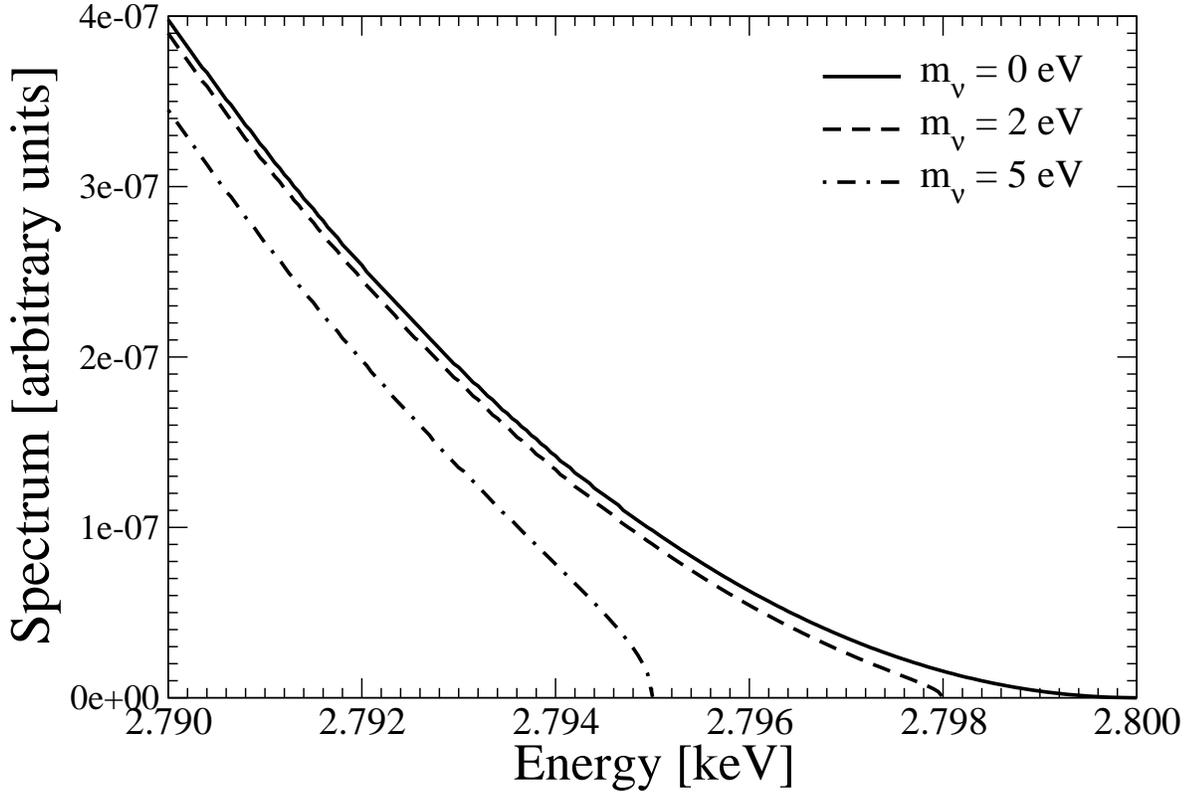}
\end{minipage}
\begin{minipage}[t]{16.5 cm}
\caption{The upper end of the bolometer (calometric) 
spectrum in the energy interval  $2.79 \  to \ 2.80 keV $ 
for the deexitation of the hole states in 
Dysprosium is shown for the neutrino masses $ m_\nu \ 
=\  0.0\  eV; \ 2.0\  eV \ and \ 5.0 \ eV $ 
and the Q 
value $Q \ =\ 2.80 \ keV $ with the overlap 
and exchange corrections $B_f$ of eq. (\ref{decay}) included 
using the values listed in table \ref{Rel}.  
\label{Spectrum279-280B}}
\end{minipage}
\end{center}
\end{figure}

\begin{figure}[tp]
\begin{center}
\begin{minipage}[t]{17cm}
\epsfig{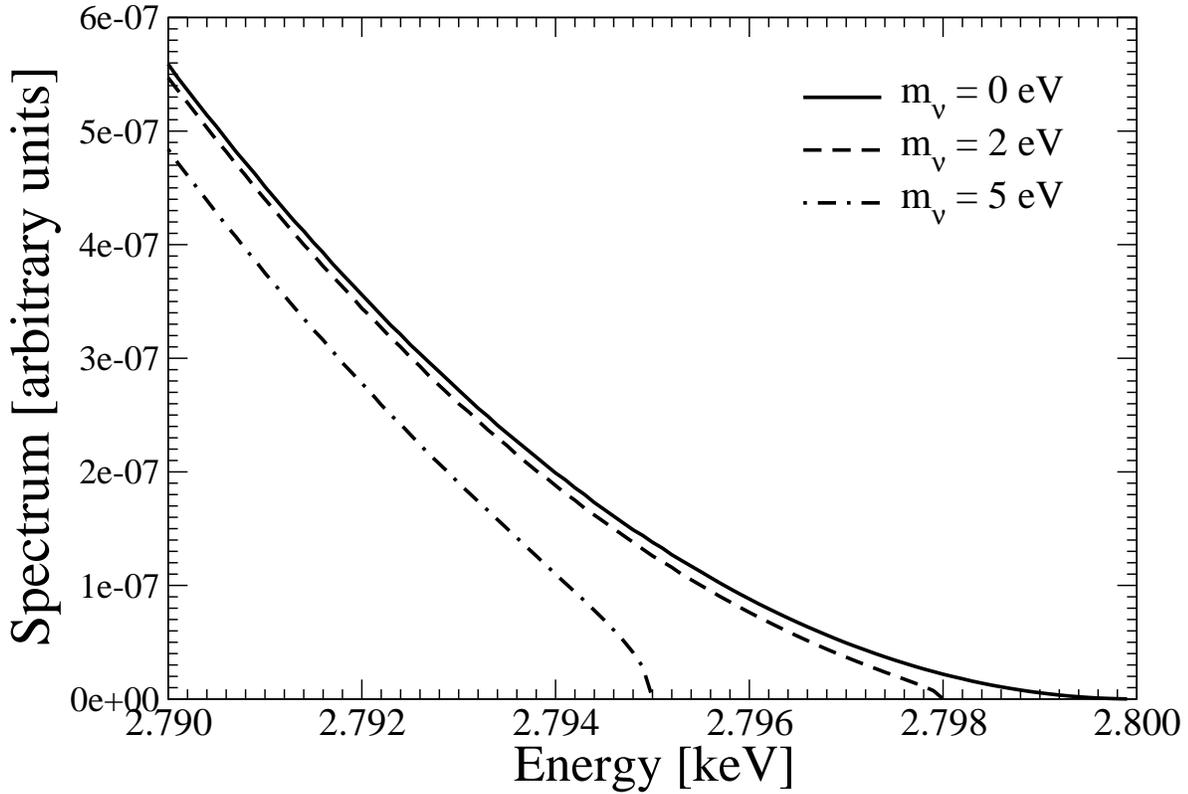}
\end{minipage}
\begin{minipage}[t]{16.5 cm}
\caption{The same bolometer (calorimetric) spectrum 
as in figure \ref{Spectrum279-280B} of the deexitation of the hole states in 
Dysprosium with the unrealistic choice of all overlap and exchange 
correction factors $B_f = 1.0$. Since then the expressions $B_f$ 
do then not show up in eq. (\ref{decay}) one can loosely speak of results  
"without overlap and exchange corrections". 
\label{Spectrum279-280}}
\end{minipage}
\end{center}
\end{figure}

\begin{figure}[tp] 
\begin{center}
\begin{minipage}[t]{17cm}
\epsfig{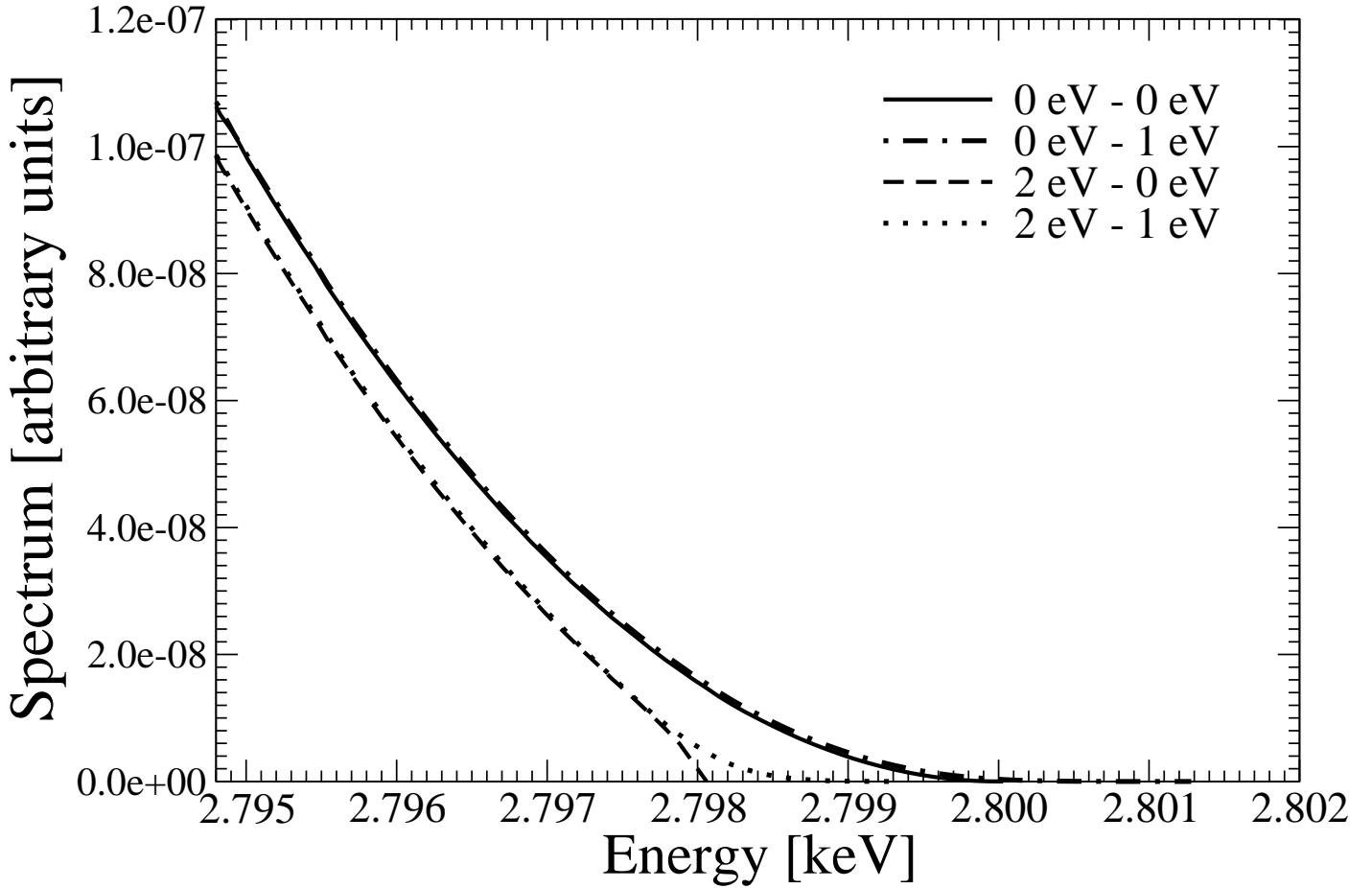}
\end{minipage}
\begin{minipage}[t]{16.5 cm}
\caption{The upper end of the bolometer (calorimetric) 
spectrum (\ref{decay}) in the energy interval  $2.7939 \  to \ 2.802 keV $ 
for electron capture in 163 Holmium 
to 163 Dysprosium for a neutrino mass of $0.0\  eV$ and $2.0\ eV$ with an energy 
resolution Full Width Half Maximum (FWHM) of $0\ eV$ and  $1\  eV$ . 
(Compare for the same neutrino masses, but different energy resolutions 
solid with dashed-dotted for $m_\nu \ = 0.0 \ eV$ and dashed 
with dotted for $m_\nu \ = \ 2.0\ eV$.) The Q 
value is assumed to be $Q \ =\ 2.80 \ keV $. Folding of the whole spectrum with a Gaussian 
to include the finite energy resolution of $ 1 \ eV$ FWHM conserves the value 
of the integral over the whole spectrum. This is not the case, if the folding 
interval is not the full spectrum. Here the folding is done over 520 
mesh points in the interval 2.7500 keV to 2.8020 keV over 52 eV.  Energy points below 
2.7500 keV do not move strength into the folding interval and energy points 
just above or equal 2.7500 keV move strength down below 2.7500 keV and this strength 
is lost for the folded spectrum.  The integral over the folded spectrum 
above 2.7500 keV is slightly by about $ 1\  \% $ (independent of the neutrino 
mass by about a factor $0.991$) smaller than the integral 
over the bare theoretical spectrum. This difference cant be seen in a figure 
and thus a renormalized curve is not shown. The overlap 
and exchange corrections $B_f$ of eq. (\ref{decay}) 
are included according to table \ref{Rel}. The calculations are performed with double 
precision to obtain the required accuracy.  
\label{Word19}}
\end{minipage}
\end{center}
\end{figure}

\begin{figure}[tp] 
\begin{center}
\begin{minipage}[t]{17cm}
\epsfig{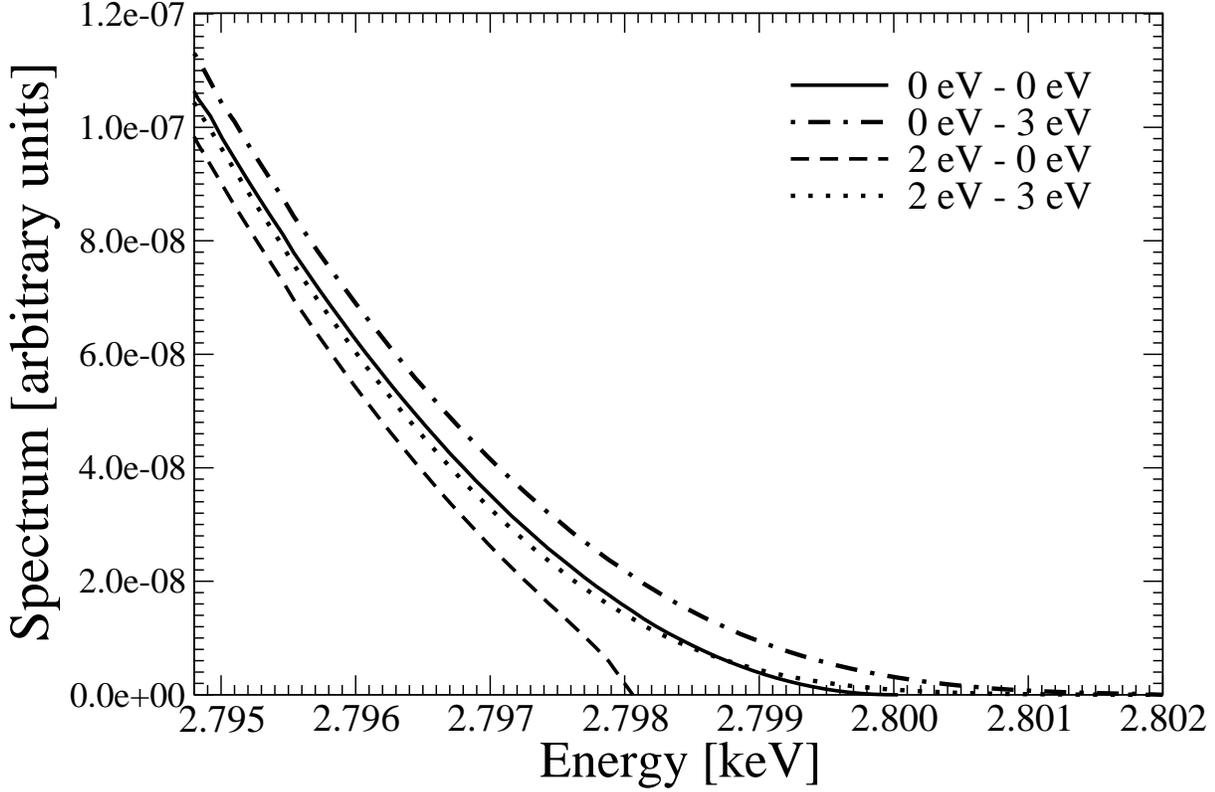}
\end{minipage}
\begin{minipage}[t]{16.5 cm}
\caption{The upper end of the bolometer (calorimetric) 
spectrum 
of electron capture in 163 Holmium 
to 163 Dysprosium (\ref{decay}) for a neutrino mass of $ 0.0 $ and $2.0 $ eV,  
with an energy resolution $ FWHM  = \  0.0 \ eV $  and $ 3.0 \ eV$. 
(Compare for the same neutrino masses and for different 
energy resolutions solid with dashed dotted for $m_\nu \ = \ 0.0\ eV$ 
and dashed with dotted for $m_\nu \ = \ 2.0\ eV$.) 
The integral over the spectrum $<2.75 \  [keV;\  2.802 \ [keV]>$ 
folded with a Gaussian with the 
Full Width Half Maximum (FWHM)  of $\ 3\  eV$ 
is slightly by about  $3\  \%$ smaller. The reasons are explained in figure \ref{Word19}. 
The exact reduction factor is by a factor three larger 
than for the folded spectrum with 1 eV FWHM from figure \ref{Word19}, 
where the reduction is only $1\ \%$.
The assumed Q value is $Q \ = \ 2.80 \ keV $. The overlap 
and exchange corrections $B_f$ of eq. (\ref{decay}) 
are included according to table \ref{Rel}. More details in the 
caption of figure \ref{Word19}. 
\label{Word18}}
\end{minipage}
\end{center}
\end{figure}

\begin{figure}[tp]
\begin{center}
\begin{minipage}[tl]{18 cm}
\epsfig{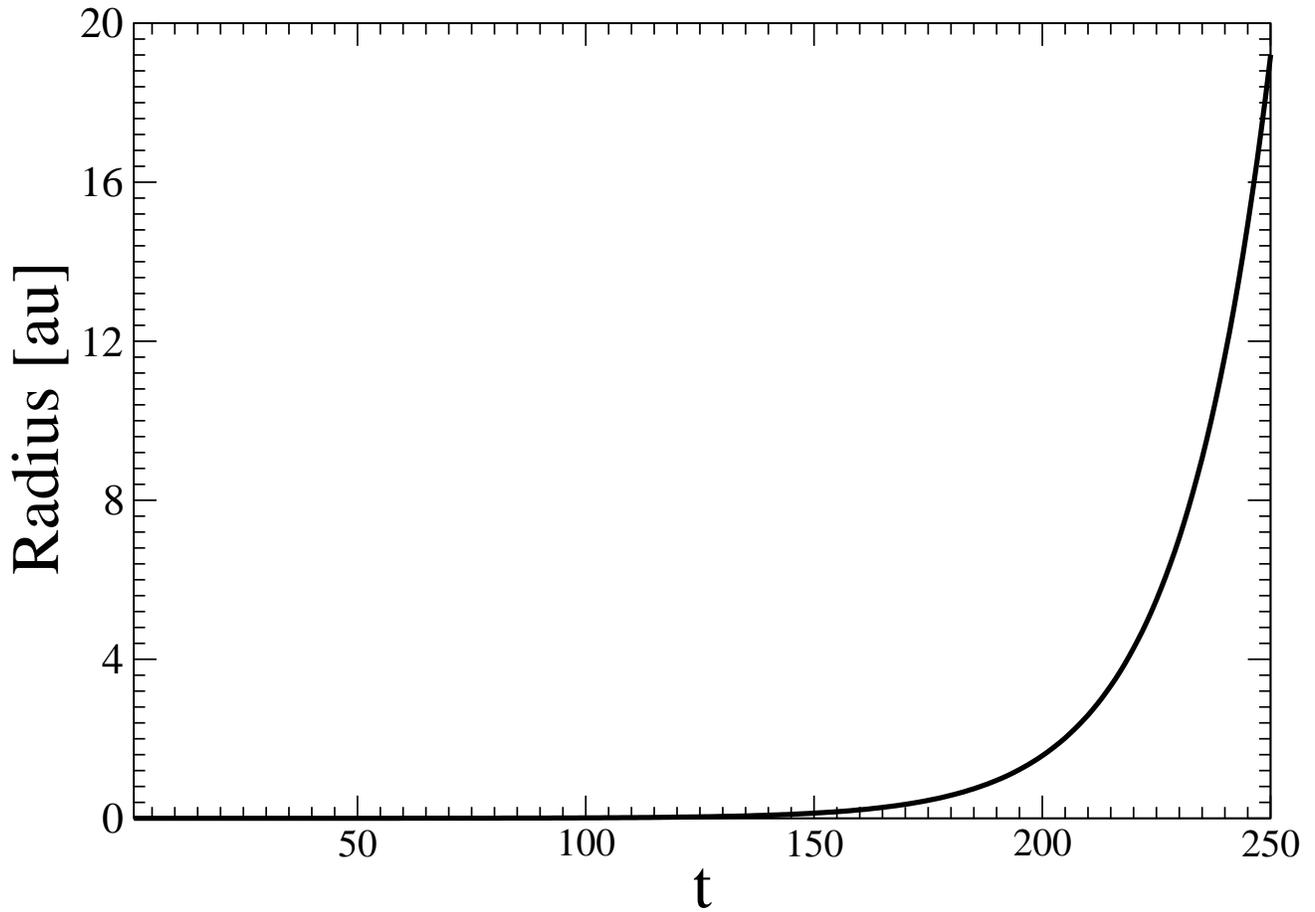}
\end{minipage}
\begin{minipage}[t]{16.5 cm}
\caption{Radial distance in a logarithmic scale $ t \  =\  ln(r/r_{0})/0.05 $, 
where t counts the radial meshpoints $ t\ = \ 1 \ to \  251$. 
 r in atomic units (Bohr radii $[a_0] $) 
for Holmium with $ r_{0} \  =  7.14693\cdot10^{-5}\ [au] $. 
\label{Radius1}}
\end{minipage}
\end{center}
\end{figure}

\begin{figure}[tp]
\begin{center}
\begin{minipage}[tl]{19 cm}
\epsfig{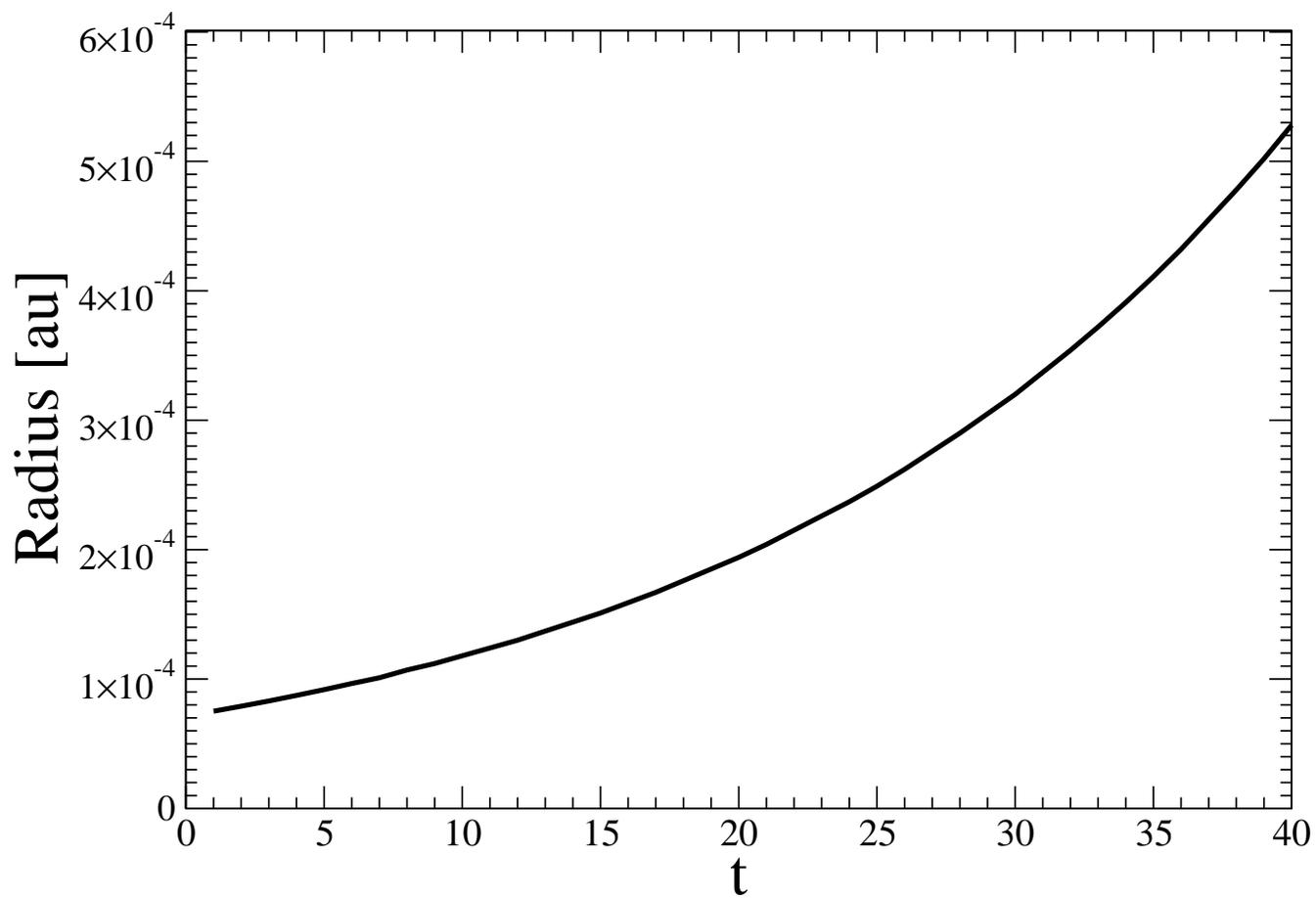}
\end{minipage}
\begin{minipage}[t]{16.5 cm}
\caption{Radial distance in Bohr radii $[a_0] $ equal to atomic units $[au]$ as a function of the 
radius parameter $ t $  used in figure $ \ref{Radius1}$ 
and in eqs. ( \ref{rr1}) and (\ref{rr2}). 
\label{Radius2}}
\end{minipage}
\end{center}
\end{figure}

\begin{figure}[tp]
\begin{center}
\begin{minipage}[t]{17 cm}
\epsfig{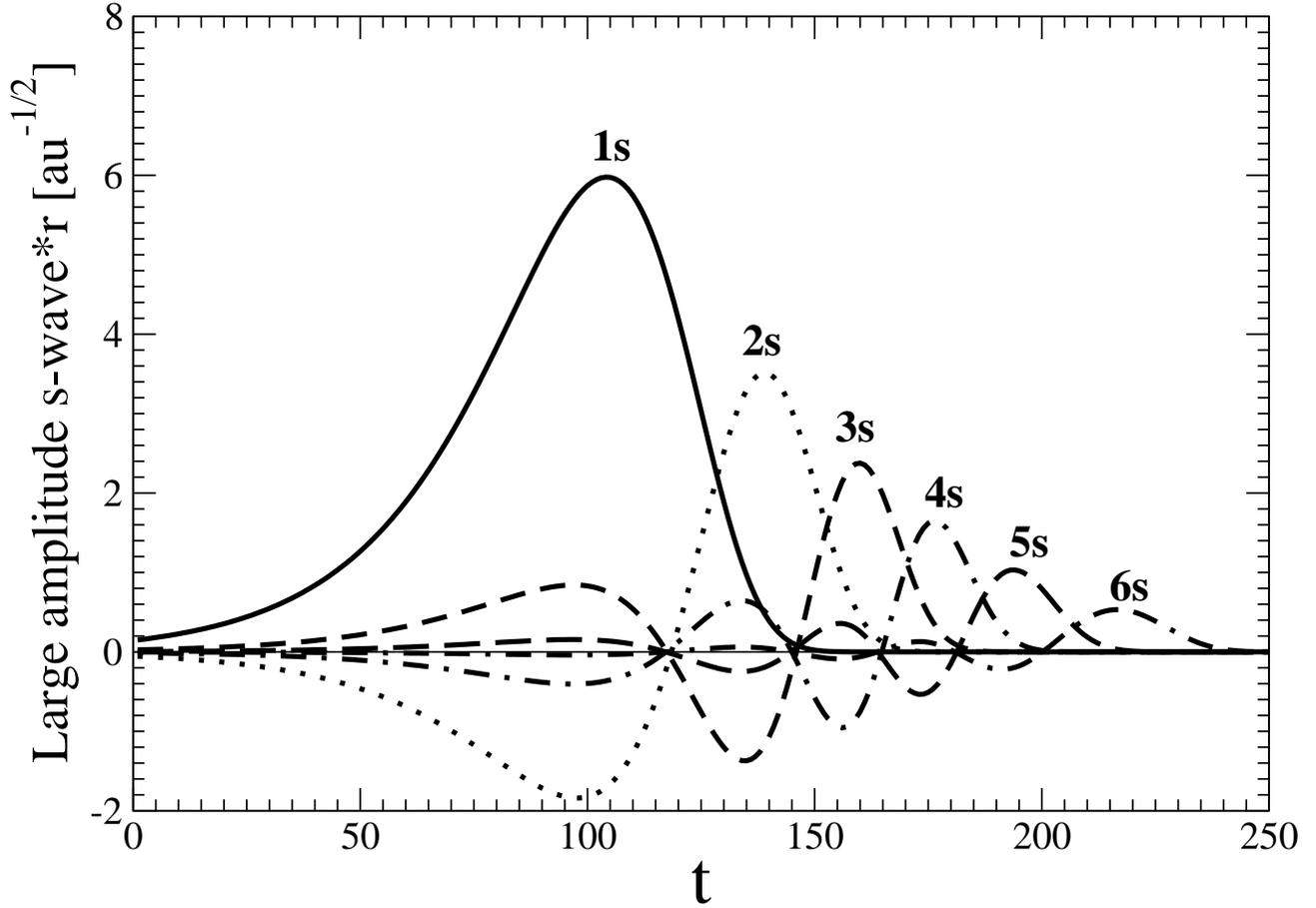}
\end{minipage}
\begin{minipage}[t]{16.5 cm}
\caption{1s (solid line), 2s (dotted), 3s (short-dashed), 4s (dashed-dotted) 
5s (long-dashed) and 6s (dashed-dashed-dotted) normalized upper spinor amplitudes of the 
electron wave functions for the Holmium ground state
multiplied by r as functions of the distance parameter t defined in
 figures (\ref{Radius1}), (\ref{Radius2}) and eqs. (\ref{rr1}) and (\ref{rr2}) . 
\label{Holmium}}
\end{minipage}
\end{center}
\end{figure}

\begin{figure}[tb]
\begin{center}
\begin{minipage}[t]{17 cm}
\epsfig{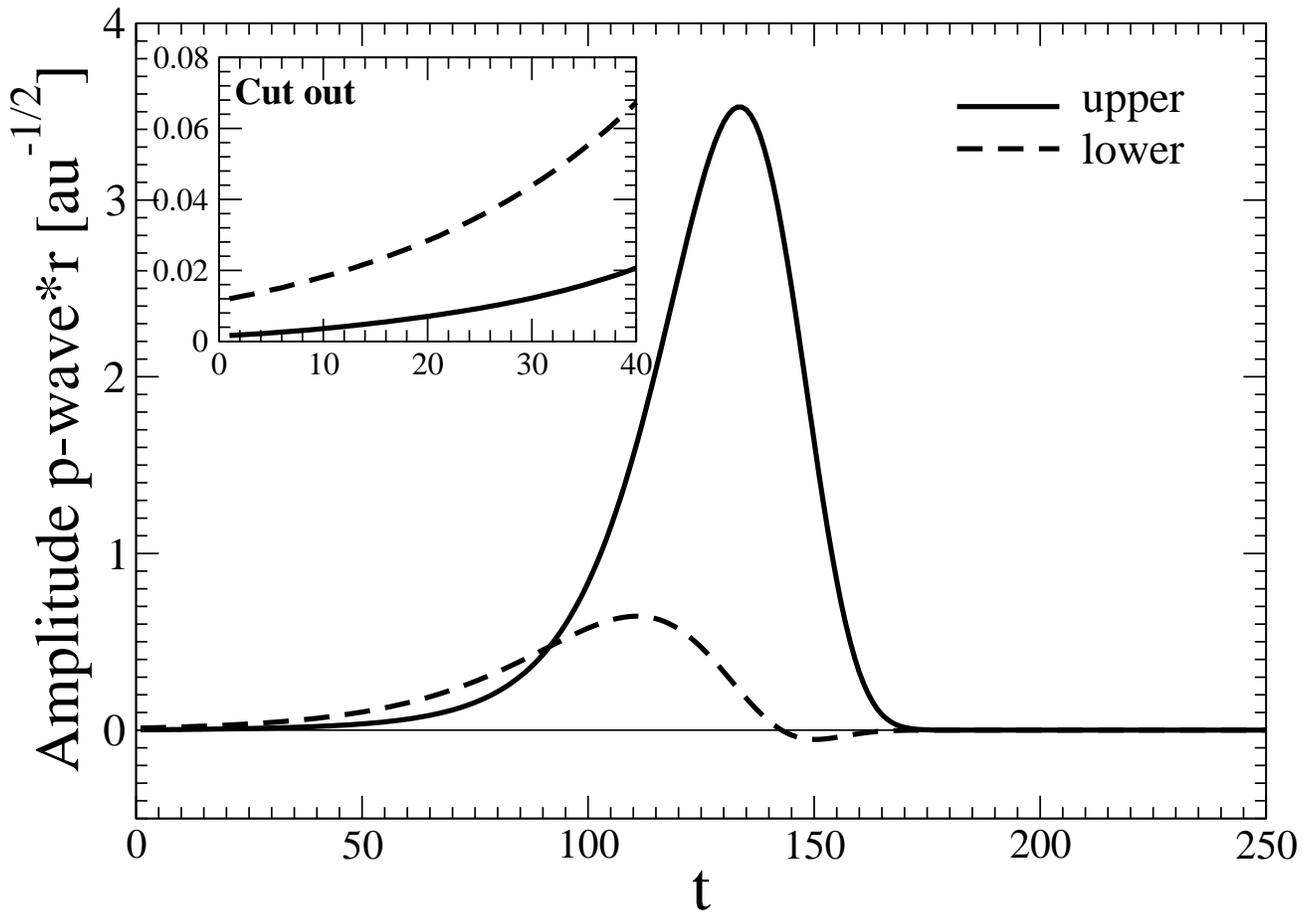}
\end{minipage}
\begin{minipage}[t]{16.5 cm}
\caption{Normalized upper (solid line) and lower (dashed) spinor amplitudes of the Ho 
electron wave functions $2p_{1/2}$ multiplied by r. The cut out shows, that the lower 
amplitude is for small r larger then the upper amplitude as a function of the logarithmic radial 
parameter t definded in eqs. (\ref{rr1}) and (\ref{rr2}) and graphically shown in figures 
(\ref{Radius1}) and (\ref{Radius2}).  
\label{smallr}}
\end{minipage}
\end{center}
\end{figure}

\begin{figure}[tb]
\begin{center}
\begin{minipage}[t]{17 cm}
\epsfig{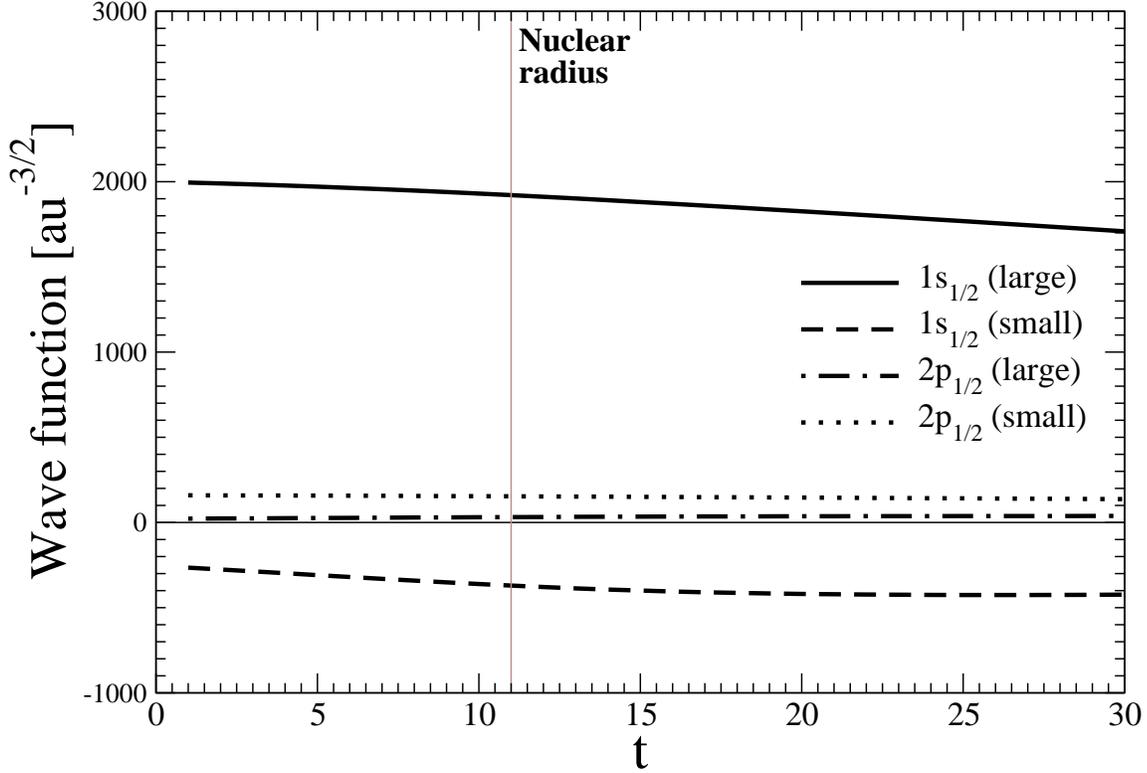}
\end{minipage}
\begin{minipage}[t]{16.5 cm}
\caption{Normalized electron wave function not multiplied by r (\ref{WF}) $P(r)/r$ 
and $Q(r)/r$ for the $ 1s$ 
and $ 2p_{1/2}$ electron in 
Holmium 163 for the ground state configuration. The figure shows, that the 
lower $2p_ {1/2}$ amplitude (here called `small') is at small r larger than the upper one. 
The logarithmic distance parameter t is defined in eqs. (\ref{rr1}) and (\ref{rr2}) 
and graphically shown in figures 
(\ref{Radius1}) and (\ref{Radius2}). The nuclear radius of the Fermi distribution 
(\ref{Ra}), (\ref{dis}) and (\ref{dif}) lies at $t = 11$ .
\label{small}}
\end{minipage}
\end{center}
\end{figure}

The main aim of this work is to study the effect of a finite 
electron neutrino mass on the calorimetric 
spectrum of the deexcitation of the hole states in $^{163}Dy$ 
after electron capture in $^{163}Ho$. A finite neutrino 
mass suppresses counts in  the  spectrum around the interval 
$(Q \ - \ m_\nu, \ Q)$. 
Since one does not know the Q 
value exactly and since the energy resolution is not perfect, 
one needs to know the theoretical 
calorimetric spectrum for extracting the neutrino mass as a function 
of the neutrino mass, the Q value and probably also of the energy resolution.
This is similar as for the  determination 
of the anti-neutrino mass in the Tritium decay of the KATRIN experiment. 
The theoretical  figures \ref{Spectrum1} 
and \ref{Spectrum2} and the experimental spectrum 
in fig. \ref{Loredana}  \cite{Gastaldo4} show, 
that the relative overlap and exchange corrections of eq. (\ref{rel}) 
are very essential for the form of the spectrum (\ref{decay}).  
From the results presented it is also clear, that 
only a fully relativistic selfconsistent treatment
can be reliable. The "small" parameter $Z/137 = 0.49$ for a non-relativistic approach 
is not so small and 
relativistic effects must be included. 
They contract the inner electron shells and increase the electron wave functions   
at the Holmium nucleus by about a factor 2 
(see table \ref{psi} ), which increases the absolute 
value of the capture cross section by a factor four. 
The relative difference between $Z$  for the parent and $(Z-1)$ 
for the daughter is getting smaller 
for heavier atoms. Thus the effect of the overlap 
and exchange corrections on the absolute value 
of the capture probability is small. But they regulate  
the relative weights of the peaks for the different hole states. 
Thus they have a large effect on 
the form of the spectrum. The finite energy resolution modifies the
 bolometer spectrum near the Q value and has to be included extracting the neutrino mass.

\vspace{1.0cm}
Acknowledgment: We want to thank members of the ECHo collaboration for discussions 
about the ECHo experiment (electron capture in Holmium 163).

\newpage

\vspace{1 cm}
\begin{table}
\caption{ 
Electron wave functions of the atomic Holmium ground state at the nuclear radius 
$ R = 6.5551 [fm] = 1.238728\cdot 10^{-4}\ [au]; \ \  
 \psi_{b,Ho}(R) \ [au^{-3/2}] $ for the upper (u) 
$P(R)/R$ and the lower $(\ell) $  $Q(R)/R$ spinor 
components of the present relativistic calculations with the Dirac-Fock 
code of Ankudinov et al.  \cite{Ankudinov} labeled by `here'. 
The non-relativistic results (FF) calculated with the 
Froese-Fischer code \ $\cite{Froese} $ \ 
are shown at the origin. Non-relativistically the p-waves disappear at $ r =   0.0 $.  
The relativistic wave functions of Mann and Waber \cite{Mann} at the origin $\psi(0)$ 
for the upper spinor components of the $ n s_{1/2}\  (\ MW \ u) $ 
and for the lower components for the \ $ n p_{1/2} (\ MW \ \ell)$ \ wave functions 
are also listed. The lower spinor component for $ n s_{1/2} $ and the upper 
component for $ np_{1/2}$  are zero at the origin. 
The increase of the wave 
functions from the non-relativistic to the relativistic approach 
at the Holmium nucleus at the nuclear radius and at the origin   
for the upper components of the $  n s_{1/2} $  
states in the relativistic approach 
by about a factor 2 is due to relativistic contractions. The phase conventions for our 
electron wave functions are different compared to the one of Mann and Waber \cite{Mann}.
}
\label{psi}
\begin{center}
\begin{tabular}{|c|r|r|r|r||c|r|r|r|} \hline
Holmium & Here u & Here $\ell$ & FF & MW u & n p1/2 & Here u 
& Here $\ell$ & MW $\ell$ \\ \hline \hline
 1s & 1769 & 425 & 1088 & 2080 & 2p1/2 &  37 & 141 & -168 \\ \hline
 2s & -648 & 157 & 359 & 763 & 3p1/2 & -18 & 70 & -82 \\ \hline
 3s & 303 & 73 & 167 & 357 & 4p1/2 & 8.8 & 33 & -39 \\ \hline
 4s & -146 & 35 & 80 & 172 & 5p1/2  &-3.1 & 11.6 & -13 \\ \hline
5s & 56 & 14 & 30 & 66 & - & - & - & -  \\ \hline
6s & -13 & 3.4 & 7.3 & 13 & - & - & - &  - \\ \hline 
\end{tabular}
\end{center}
\end{table}

\vspace{1 cm}

\begin{table}
\caption{ Overlaps (contractions of Wick's theorem) of the relativistic single 
electron wave functions of the excited Z = 66 Dysprosium* 
with a vacancy in the 3s state  and Z = 67 Holmium in the ground 
state for $ns_{1/2}$ , $np_{1/2}$ and $np_{3/2} $ for  
 $ <n' \ell', Dy(3s)^{-1}| n\ell,Ho> $.}
\label{Dy3s}
\begin{center}

\begin{tabular}{|c |r|r|r|r|r|r|r|r|} \hline
3s Dy hole & 1s & 2s      & 3s       & 4s       & 5s       & 6s        & - & - \\ \hline 
\hline
1s' & 0.999910 & 0.008148 &-0.003119 & 0.001433 &-0.000546 &  0.000135 & - & - \\ \hline
2s' &-0.007968 & 0.999716 & 0.015560 &-0.005604 & 0.002047 & -0.000503 & - & -  \\ \hline
3s' & 0.003085 &-0.015116 & 0.999389 & 0.023618 &-0.006894 &  0.001342 & - & - \\ \hline
4s' &-0.001447 & 0.005660 &-0.023007 & 0.999332 & 0.013934 & -0.003089 & - & - \\ \hline
5s' & 0.000547 &-0.002058 & 0.006868 &-0.013227 & 0.999510 &  0.010002 & - & - \\ \hline
6s' &-0.000135 & 0.000506 &-0.001653 & 0.002964 &-0.009576 &  0.999783 & - & - \\ \hline 
\hline
 -  &$2p_{1/2}$ & $3p_{1/2}$ & $4p_{1/2}$ & $5p_{1/2}$ & $2p_{3/2}$& 
$3p{3/2}$ & $4p_{3/2}$& $5p_{3/2}$  
\\ \hline \hline
$2p'_{1/2}$ & 0.999801 & 0.014854 &-0.005069 & 0.001685 &  -  & - & - & -\\ \hline
$3p'_{1/2}$ &-0.013756 & 0.999563 & 0.016686 &-0.004762 &  -  & - & - & -\\ \hline
$4p'_{1/2}$ & 0.005058 &-0.016148 & 0.999524 & 0.011956 &  -  & - & - & - \\ \hline
$5p'_{1/2}$ &-0.001671 & 0.004639 &-0.011410 & 0.999594 &  -  & - & - & -  \\ \hline
$2p'_{3/2}$ &   - &     - &   - &   - & 0.999846 & 0.012517 &-0.004483 & 0.001435 \\ \hline
$3p'_{3/2}$ &   - &     - &   - &   - &-0.012259 & 0.999648 & 0.014995 &-0.004144 \\ \hline
$4p'_{3/2}$ &   - &     - &   - &   - & 0.004481 &-0.014573 & 0.999625 & 0.010227 \\ \hline 
$5p'_{3/2}$ &   - &     - &   - &   - &-0.001428 & 0.004052 &-0.009814 & 0.999707 \\ 
\hline \hline 
\end{tabular}
\end{center}
\end{table}

\vspace{1 cm}

\begin{table}
\caption{ The first column gives the empty electron states in the 
daughter atom (Dysprosium). The approximate 
amplitudes $AF_{0}$ \ (\ref{Approx0}) and $AF$ (\ref{Approx}) and also $AFM_{0}$ and $AFM$ 
\cite{Mann} are calculated 
assuming, that the overlaps $<k'|k>\   =\ 1.0 $ 
with the same quantum numbers in the parent and the daughter are unity 
(Vatai approximation \cite{Vatai, Vatai2}).  
The amplitudes $F_{0}$  in eq. (\ref{F0f}) and $FM_{0}$, include  the full overlap corrections 
but no exchange terms. In this approximation the results in the column $ AF_0 $ 
are given by the upper amplitude (\ref{WF}) as $ P(R)/R$ 
at the nuclear radius for the s-states and by the lower amplitude as $ Q(R)/R $ 
for the $ p_{1/2}$-states (see table \ref{psi}). $F_0$ includes the overlap corrections.  
Thus $F_0$ is always in its absolute 
value smaller than $AF_0$. The exchange contributions are then in addition included 
in $F$ eq. (\ref{Ff}) and in $FM$. All quantities with M are calculated with the tabulated 
relativistic electron wave functions $ns_{1/2}$ and $np_{1/2}$ of Mann and Waber \cite{Mann} 
at the origin $\psi(0)$ with overlaps of this work with relativistic 
wave functions of the Ankudinov code \cite{Ankudinov}. 
(See table \ref{psi}). The relative sign of  wave functions, which mix  
(same $\ell$ and j but different n) are important. 
But a common minus sign of wave function amplitudes 
 is irrelevant, even if they mix.  To be consistent in the sign choice between our overlaps
 and the wave functions of Mann and Waber \cite{Mann} one has 
to adjust the phase conventions to each other.}

\label{FNS}
\begin{center}

\begin{tabular}{|c |r|r||r|r||r|r||r|r|} \hline
 -         & $ AF_{0}$ & $AF$    & $ F_{0}$& $F$    
&$AFM_{0}$& $AFM$&$FM_{0}$&$FM$ \\	\hline \hline
$1s$       & 1769.2    & 1762.7  & 1767.8 & 1761.3 &  2080   & 2072 &  2078   & 2071\\ \hline
$2s$       & -648.6    &  -657.4 & -645.1  & -653.8 &  -763   & -773 &  -759   & -769\\ \hline
$3s$       & 303.2     &  315.1  &  298.2  & 309.9  &   357   &  371 &   351   &  365\\ \hline
$4s$       & -146.3    & -157.4  & -142.4  & -153.2 &  -172   & -185 &  -168   & -180\\ \hline
$5s$       & 56.3      &  63.3   &  53.3   & 55.5   
&    66.3 & 74.6 &    62.7   &   65.3\\ \hline \hline
$2p_{1/2}$ & 141.6     &  140.3  & 140.9  & 139.7 & 166.2  & 167.3& 165.4 &  166.5\\ \hline
$3p_{1/2}$ & -69.8     &  -71.0  & -68.7   & -69.9  &    -82  &   -80&    -80  &  -79\\ \hline
$4p_{1/2}$ &  33.2     &   35.1  &  32.3   &  34.2  &   38.9  &  38.3&   37.8  &  37.3\\ \hline
$5p_{1/2}$ & -11.6     &  -13.0  & -11.0  & -12.3  
&  -13.6  & -12.9&  -12.8 & -12.2\\ \hline \hline
\end{tabular}
\end{center}
\end{table}

\vspace{1 cm}

\begin{table}
\caption{ Relative probabilities (\ref{rel}) of capture 
from $Z \ =\ 67, \ A \ = \  163\ $  Holmium to 
different states in $ Z = 66, \  A \ = \  163\ $ Dysprosium. 
Due to the small Q value of around $2.5\  keV $ 
only vacancies in $ M_1(3s_{1/2})$ and $M_2(3p_{1/2}) $ and higher 
can be excited due to energy conservation. (See also (\ref{BE}) ). 
The ratios are $ N_1/M_1, O_1/M_1, M_2/M_1, N_2/M_1 $ and $ O_2/M_1 $. $ABr_0$ 
is the approximate expression without 
exchange terms calculated with $AF_0$ (\ref{Approx0}) and $ABr$ 
with the approximate expression $AF$ (\ref{Approx}).
The ratios $Br_0 $ and $Br$ are the full expression without 
exchange and with exchange terms according to eqs. (\ref{F0f})
and (\ref{Ff}) for the ratios given above. The quantities with M give the equivalent 
results using the relativistic wave functions at the origin tabulated 
by Mann and Waber and given in table (\ref{psi}) \cite{Mann} 
and the overlaps calculated with our 
relativistic electron wave functions. (See for example table \ref{Dy3s}.) } 
The $B_f$ shown here in column 5 should be used in eq. (\ref{decay}) 
to calculate the (calometric) bolometer spectrum in arbitrary units 
to compare with the data of figure \ref{Loredana}.

\label{Rel}
\begin{center}

\begin{tabular}{|c |r|r|r|r|r|r|r|r|} \hline
 -         &  $ ABr_0$  & $ABr $   &   $ Br_0$ & $ Br$ &$ABr_{0}M$&$ABrM$&$Br_{0}M$&$BrM$ \\ 
\hline \hline
$4s$       & 0.233     & 0.250   & 0.228    & 0.244&  0.233  &0.249&  0.228 &0.244 \\ \hline
$5s$       & 0.034     & 0.040   & 0.032    & 0.032&  0.034  &0.038&  0.032 &0.032 \\ \hline
$3p_{1/2}$ & 0.053     & 0.051   & 0.053    & 0.051&  0.052  &0.046&  0.052 &0.047 \\ \hline
$4p_{1/2}$ & 0.012     & 0.012   & 0.012    & 0.012&  0.012  &0.011&  0.012 &0.010 \\ \hline
$5p_{1/2}$ & 0.001     & 0.002   & 0.001    & 0.002&  0.001  &0.001&  0.0013 &0.001 \\ \hline 
\hline

\end{tabular}
\end{center}
\end{table}

\begin{table}
\caption{ The relative intensity of electron capture from different orbitals in 
Holmium relative to capture from the $3s_{1/2}, \ M_1$ orbit . 
The finite radius of the nucleus allows also to include capture 
from other than $ns_{1/2}$ and  $np_{1/2}$ electrons. At the radius also other orbitals 
have wave functions different from zero.  The relative intensities are calculated 
in the so called  Vatai \cite{Vatai, Vatai2} 
approximation (\ref{RU}).  The first column gives the quantum 
number of the denominator, which is always the $M_1$ 
orbit $3s_{1/2}$.  The quantum numbers in the rows characterize the wave functions for   
the numerators (\ref{RU}), for which the relative intensity is given. 
The $"\ell"$ indicates,
that for the $j \, = \, \ell \,  - \, 1/2$ states the larger lower component of the Dirac wave function is used at the nuclear radius. 
Capture from other than $s_{1/2}$ and $p_{1/2}$ states are 
extremely small and can be neglected.   
}

\label{Vat}
\begin{center}

\begin{tabular}{|c ||r|r|r|r|r|} \hline 
 -         & $ 3s_{1/2}$ & $3p_{1/2}\ell $&$3p_{3/2}$& $3d_{3/2}\ell$ & $3d_{5/2}$ \\	
\hline 
$3s_{1/2}$ & 1.000 & 0.048  & $3\cdot10^{-6}$& $2\cdot10^{-8}$ & $5\cdot10^{-13}$\\ 
\hline \hline
-      &$4s_{1/2}$ & $4p_{1/2}\ell$ & $4p_{3/2}$ & $4d_{3/2}\ell$ & $4d_{5/2}$\\ \hline
$3s_{1/2}$ & 0.233 & 0.012 & $ 7\cdot10^{-8}$ & $4\cdot10^{-9}$ & $1\cdot10^{-13}$\\ 
\hline \hline 
-      & $4f_{5/2}\ell$ & $4f_{7/2}$ & $5s_{1/2}$ &$ 5p_{1/2}\ell$& $5p_{3/2}$\\ \hline 
$3s_{1/2}$  & $ 4\cdot10^{-17}$  & $7\cdot10^{-22}$ & 0.034  & 0.002   
& $8\cdot10^{-8}$\\ \hline  
\end{tabular}
\end{center}
\end{table}

\newpage 

\vspace{1cm}

\vspace{1cm}

\vspace{1cm}

\end{document}